\documentclass[10pt]{article}
\usepackage{graphics}
\usepackage{graphicx}

\begin{document}
%\draft

% below commands are for enumerating eqs according to sections
%\newcounter{eq}[section]
%\newcommand{\set}{\stepcounter{eq}
%\renewcommand{\theequation}{\mbox{\arabic{section}.\arabic{eq}}}}

% command for enumerating Refs with 1. instead of [1]:
\makeatletter
\renewcommand{\@biblabel}[1]{\quad#1.}
\makeatother

\hsize=6.15in
\vsize=8.2in
\hoffset=-0.42in
\voffset=-0.3435in

\normalbaselineskip=24pt\normalbaselines

\begin{center}
  {\Large \bf Cooperativity, information gain, and energy cost during early LTP 
    in dendritic spines }
\end{center}

\vspace{0.15cm}

\begin{center}
  { Jan Karbowski$^{1,*}$,
    Paulina Urban$^{2,3,4}$} 
\end{center}

\vspace{0.05cm}

\begin{center}
 {\it $^{1}$ Institute of Applied Mathematics and Mechanics, University of Warsaw, Warsaw, Poland;
  $^{2}$ College of Inter-Faculty Individual Studies in Mathematics and Natural Sciences,
 University of Warsaw, Warsaw, Poland;  $^{3}$ Laboratory of Functional and Structural Genomics,
 Centre of New Technologies, University of Warsaw, Warsaw, Poland; $^{4}$ Laboratory of Databases
 and Business Analytics, National Information Processing Institute, National Research Institute,
 Warsaw, Poland. }
\end{center}

%\date{\today}

\vspace{0.1cm}

%\widetext
\begin{abstract}
We investigate a mutual relationship between information and energy during early phase of LTP
induction and maintenance in a large-scale system of mutually coupled dendritic spines, with
discrete internal states and probabilistic dynamics, within the framework of nonequilibrium
stochastic thermodynamics. In order to analyze this computationally
intractable stochastic multidimensional system, we introduce a pair approximation, which
allows us to reduce the spine dynamics into a lower dimensional manageable system of closed
equations. It is found that the rates of information gain and energy attain their maximal
values during an initial period of LTP (i.e. during stimulation), and after that they recover
to their baseline low values, as opposed to a memory trace that lasts much longer. This
suggests that learning phase is much more energy demanding than the memory phase. We show
that positive correlations between neighboring spines increase both a duration of memory
trace and energy cost during LTP, but the memory time per invested energy increases
dramatically for very strong positive synaptic cooperativity, suggesting a beneficial
role of synaptic clustering on memory duration. In contrast, information gain after LTP
is the largest for negative correlations, and energy efficiency of that information generally
declines with increasing synaptic cooperativity. We also find that dendritic spines can use
sparse representations for encoding of long-term information, as both energetic and structural
efficiencies of retained information and its lifetime exhibit maxima for low fractions of
stimulated synapses during LTP. Moreover, we find that such efficiencies drop significantly
with increasing the number of spines. In general, our stochastic thermodynamics approach
provides a unifying framework for studying, from first principles, information encoding and
its energy cost during learning and memory in stochastic systems of interacting synapses.

\end{abstract}

%\pacs{PACS Nos. 87.18.Sn, 87.19.La, 87.80.Xa, 89.40.+k, 87.23.Ge}
%\narrowtext 
%\maketitle 

%\begin{narrowtext}

%\maketitle


\vspace{0.1cm}

\noindent $^{*}$ Corresponding author:  jkarbowski@mimuw.edu.pl

%\noindent Phone: (626)-395-5840, Fax: (626)-795-2397.

\vspace{2.3cm}

%\newpage

%\noindent {\bf Author Summary:}

%\vspace{0.4cm}

%\newpage

%{\large \bf Significance statement}

\newpage

{\Large \bf 1. Introduction.}

Parts of synapses known as dendritic spines play an important role in learning and memory
in neural networks (Bonhoeffer and Yuste 2002; Kasai et al 2003; Bourne and Harris 2008;
Takeuchi et al 2014; Kandel et al 2014). Learning can be thought as acquiring information in
synapses through plasticity mechanism such as LTP and LTD (long term potentiation and depression,
respectively), and memory can be regarded as storing that information (for experimental overview
see e.g.: Yang et al 2009; Bourne and Harris 2008; Takeuchi et al 2014; Poo et al 2016; for
theoretical overview see e.g.: Fusi et al 2005; Benna and Fusi 2016; Chaudhuri and Fiete 2016).
Thus learning and memory are strictly related to processing and maintaining of long
term information, which in principle could be quantified in terms of information theory
and statistical mechanics, similar as it was done for neural spiking activity (Rieke et al
1999). Indeed, recent results by the authors show that information (entropy) contained in
the distributions of dendritic spine volumes and areas is nearly maximal for any given of
their average sizes across different brains and cerebral regions (Karbowski and Urban 2022).
This suggests that the concept of information can be useful in quantifying synaptic learning
and memory, and that actual synapses might ``use'' and optimize certain information-theoretic
quantities.

Physics teaches us that there are close relationships between information and thermodynamics,
and the smaller the system the stronger the mutual link, since smallness enhances fluctuations
and thus unpredictability in the system (Bennett 1982; Leff and Rex 1990; Berut et al 2012;
Parrondo et al 2015). This means that information processing always requires some energy,
and it is reasonable to assume that biological evolution favors systems that save energy
while handling information, because of the limited resources and/or competition (Niven and
Laughlin 2008). This line of thought was explored in neuroscience for estimating energy-efficient
coding capacity in (short-term) neural activities and synaptic transmissions (Rieke et al 1999;
Levy and Baxter 1996 and 2002; Levy and Calvert 2021; Laughlin et al 1998; Balasubramanian
et al 2001). All these approaches and calculations for neural and synaptic activities, however
valuable, missed one key ingredient of real biological systems, or did not make it explicit.
Namely, biological systems, including neurons and synapses, always operate far from
thermodynamic equilibrium, where balance between incoming and outgoing energy and matter
(or probability) fluxes is broken; the so-called broken detailed balance (for a general
physical approach, see: Maes et al 2000; Seiferet 2012; Gardiner 2004; for a biophysical
approach to synapses, see: Karbowski 2019). As a consequence, equilibrium thermodynamics
(with static or stationary variables and Gibbs distributions) does
not seem to be the right approach, and to be more realistic, one has to use nonequilibrium
statistical mechanics, where the concepts of stochasticity and entropy production play central
roles (and where we do not know a priori the probability distributions). Nonequilibrium
statistical mechanics (or nonequilibrium stochastic thermodynamics) provides a unifying
description for stochastic dynamics, because it treats microscopic information and energy on the
same footing, which allows us to get the right estimates of information and energy rates from
first principles. Such an approach was initiated in (Karbowski 2019, 2021) for studying
nonequilibrium thermodynamics of synaptic plasticity. Both of these studies suggested that
synaptic plasticity can use energy rather economically, since (i) it consumes only about
$4-11 \%$ of energy devoted for fast synaptic transmission (Karbowski 2019), and (ii) it can
provide higher coding accuracy and longer memory time for a lower energy cost at certain regimes
(Karbowski 2019, 2021).

There exists a large experimental evidence that local synaptic cooperativity on a dendrite
takes place during learning and memory formation (Marino and Malinow 2011; Yadav et al 2012;
for a review see: Winnubst et al 2012), and it can also be useful for long-term memory stability
(Govindarajan et al 2006; Kastellakis and Poirazi 2019). Thus, it seems that any realistic
model of synaptic plasticity relevant for memory formation should include correlations
between neighboring synapses. Moreover, it would be good to know how such correlations
influence the lifetime of memory trace, as well as, information gain and energy cost
associated with it.

The current study explores nonequilibrium statistical mechanics for investigating the
efficiency of learning and storing information in a system of interacting synapses with
stochastic dynamics during early LTP induction and its maintenance (e-LTP phase, without
consolidation). Not only fast synaptic transmission is noisy (Volgushev et al 2004); the noise
is also present in long-term synaptic dynamics associated with slow plasticity due to large
thermal fluctuations in internal molecules and presynaptic input (Bonhoeffer and Yuste 2002;
Holtmaat et al 2005; Choquet and Triller 2013; Statman et al 2014; Meyer et al 2014;
Kasai et al 2003; Loewenstein et al 2011). Thus synaptic plasticity requires a probabilistic
approach (Yasumatsu et al 2008), and we assume that it can be described as dynamics involving
transitions between discrete mesoscopic states (e.g. Montgomery and Madison 2004; Fusi et al 2005;
Leibold and Kempter 2008; Barrett et al 2009; Benna and Fusi 2016). Our paper extends the
previous two studies (Karbowski 2019, 2021) in three important methodological ways.
First, it provides a general framework for approximating the dynamics of
multidimensional stochastic system of $N$ interacting synapses each with 4 internal
states (which in practice is computationally intractable for large $N$),
by reducing it to a set of coupled lower-dimensional stochastic subsystems (which are
computationally tractable). This is done by applying the so-called pair approximation, which
allows us to reduce the system with $4^{N}$ equations to the system with $\sim 4(5N-4)$ equations.
Second, our paper provides explicit formulas for investigating information gain (rate of
Kullback-Leibler divergence) and energy consumption (entropy production rate) for arbitrary
time during learning and memory retention phases. Third, we use data-driven estimates for
transition rates between different synaptic states, and hence our values of information and
energy are realistic.

On a conceptual level, our study investigates how the efficiency of encoded information
(both amount and duration) depends on a degree of correlation between neighboring synapses,
on a percentage of their activation by presynaptic neurons, and on magnitude and duration of
synaptic stimulation during LTP. The first relates to cooperativity between synapses, the second
to sparseness of synaptic coding, and the last to the strength of learning. All these parameters
can in principle be compared to empirical data, once those are available, thus providing an
important link between theory and experiment.

Throughout the paper we use interchangeably the terms ``dendritic spine'' and ``synapse''.

\newpage

{\Large \bf 2. Model of plastic interacting dendritic spines.}

We consider a single postsynaptic neuron having one basal (main) dendrite
with $N$ dendritic spines located along its length (Fig. 1). Because of
the small sizes of dendritic spines ($\sim 1 \mu$m) their dynamics is necessarily
probabilistic due to thermodynamic fluctuations of local environment (with high
temperature $\sim$ 300 K), as well as due to activity fluctuations in neurons
(both of electric and chemical nature).
Moreover and importantly, the spines are locally coupled by nearest neighbor
interactions, as the experimental data suggest (Marino and Malinow 2011; Yadav
et al 2012; Winnubst et al 2012)). 
 
\vspace{0.4cm}

\noindent{\large \bf 2.1. Morphological spine states and stochastic multidimensional dynamics.}

Empirical data indicate that a single dendritic spine can be regarded as a four state
stochastic system with well defined morphological (mesoscopic) states (Montgomery and Madison
2004; Bokota et al 2016; Basu et al 2018; Urban et al 2020; Fig. 1A). These states are denoted
here as $s_{i}$ at each location $i$, with values $s_{i}= 0, 1, 2, 3$, corresponding respectively
to the following morphological states: nonexistent, stubby, filopodia/thin, and mushroom
(the larger $s_{i}$, the larger the spine size; see Appendix A). These mesoscopic states are
quasi stable, which means that there are slow transitions between them that are much slower
than microscopic transitions between molecular, mostly unknown, processes comprising internal
microscopic dynamics of a dendritic spine (Kennedy 2000; Sheng and Hoogenraad 2007;
Miller et al 2005; Kandel et al 2014). This approach can be treated as a coarse-grained
description of intrinsic spine dynamics in terms of stochastic Markov process on a mesoscopic
scale.

We assume stochastic dynamics for $N$ coupled dendritic spines, and denote by
$P(s_{1}, s_{2}, ..., s_{N};t)$ the probability that the spine system has the
configuration of internal states $s_{1}, s_{2}, ...., s_{N}$.
The dynamic of this global stochastic state is motivated
by the Glauber model of time-dependent Ising model (Glauber 1963), and it is
represented by the following Master equation (see Appendix A)

\begin{eqnarray}
\frac{d P(s_{1},...,s_{N})}{dt}= \sum_{i=1}^{N} \sum_{s'_{i}} w_{s_{i},s'_{i}}(s_{i-1},s_{i+1})
 P(s_{1},...,s'_{i},...,s_{N})    \nonumber  \\ 
  - P(s_{1},...,s_{N}) \sum_{i=1}^{N} \sum_{s'_{i}} w_{s'_{i},s_{i}}(s_{i-1},s_{i+1}),
\end{eqnarray}\\
where $w_{s_{i},s'_{i}}(s_{i-1},s_{i+1})$ is the transition rate for the jumps inside
spine $i$ from state $s'_{i}$ to state $s_{i}$. These jumps also depend on the states
of neighboring spines $s_{i-1}$ and $s_{i+1}$, because of the nearest neighbor coupling
between the spines (Fig. 1B). Generally, we take the following form of the transition matrix
$w_{s_{i},s'_{i}}(s_{i-1},s_{i+1})$:

\begin{eqnarray}
  w_{s_{i},s'_{i}}(s_{i-1},s_{i+1}) = v_{s_{i},s'_{i}}
  \left[1 + \theta(d(s_{i})-d(s'_{i}))\frac{\gamma}{2d(3)}[d(s_{i-1})+d(s_{i+1})]\right]
\nonumber  \\   
   \times \left[1 + a_{i}(1+\theta(d(s_{i})-d(s'_{i})))f(t)\right]  
\end{eqnarray}\\ 
where $v_{s_{i},s'_{i}}$ is the intrinsic basic transition rate between $s'_{i}$ and $s_{i}$ at spine
$i$, and it is independent of the neighboring synapses. These intrinsic rates are the same for each
$i$, setting the temporal scale for basal synaptic plasticity, and they were estimated based on
data in (Urban et al 2020) and are presented in Table 1. The symbol $d(s_{i})$
is the spine size at state $s_{i}$ of spine $i$ and it is proportional to spine surface area
(see Appendix A), whereas $\theta(x)$ denotes the sign function of the argument $x$, where
$\theta(x)= 1$ if $x \ge 0$, and $\theta(x)= -1$ if $x < 0$. Note that larger spine sizes of
neighboring spines generally influence more the transition rates of the given spine, which relates
to their cooperativity. The parameter $\gamma$ corresponds to magnitude of spine cooperativity
between nearest neighbors, with $-1 \le \gamma \le 1$, and there is rescaling by the maximal
spine size $d(3)$ (size in state $s=3$ called mushroom), which ensures positivity of all
elements of the transition matrix.
The positive values of $\gamma$ indicate positive correlations (positive cooperativity),
while its negative values mean negative correlations (negative cooperativity). Note that for
positive cooperativity, the local spine interactions amplify the transitions that lead to increase
of spine size, and reduce those transitions that decrease spine size. The opposite is true for
negative cooperativity. It should be added that distant spines also can affect a given spine at
location $i$, but that interaction has an indirect character and thus is weaker and mediated
with some delay.

The last factor on the right in Eq. (2) indicates the effect of external (presynaptic) stimulation,
leading to LTP, with a time varying function known as alpha function $f(t)$ given by

\begin{eqnarray}
f(t)= A(e^{-t/\tau_{1}} - e^{-t/\tau_{2}}),
\end{eqnarray}\\ 
where $A$ is the stimulation amplitude, $t$ is the time after stimulation onset, and
$\tau_{1}, \tau_{2}$ are time constants related to falling and rising phases of the stimulation,
respectively. The latter means that LTP related stimulation last only about $\tau_{1}+\tau_{2}$
($\sim 17$ min; Table 2), which we call the duration of the learning phase. After that time,
all the transition rates essentially decay to their pre-stimulation basal values. Consequently, 
after the stimulation, the dynamics of synaptic plasticity is driven only by the interactions
between neighboring spines and their internal states. These dynamics are slow, and we call
this stage the memory phase. The prefactor $a_{i}$ of $f(t)$ in Eq. (2) assumes two values:
$a_{i}=1$ when the spine $i$ is stimulated with the probability $p_{act}$,
and $a_{i}=0$ when there is no stimulation with the probability $1-p_{act}$.
Note that LTP related stimulation amplifies only the transitions increasing the spine size
(the sign function $\theta$ is then 1). For the transitions decreasing the spine size, there is
no amplification, because then the prefactor of $f(t)$ is zero.

The important point is that during learning (the phase when the function $f(t)$ is activated),
the information about the stimulation is encoded in the patterns of the probabilities
$P(s_{1},...,s_{N})$. Thus, knowing how these patterns change in time provides necessary
``data'' for computing physical characteristics of learning and memory.

\vspace{0.4cm}

\noindent{\large \bf 2.2. Reduction of multidimensional spine stochastic dynamics
  into low dimensional dynamics: pair approximation.}

Equation (1) describes dynamics of the multidimensional probability that involves a gigantic
$4^{N}$ number of equations. For a typical number of synapses on a dendrite $N\sim 10^{3}$,
the dynamics represented by Eq. (1) require $\sim 10^{600}$ equations, which are impossible
to simulate and solve on any existing computer. This numerical impossibility forces us to find
an approximation to the dynamics in Eq. (1). Consequently, we consider a lower dimensional
dynamics involving only singlets and pairs of locally interacting spines. This strategy is
sufficient to describe the global dynamics of the synaptic system, and to compute information
and energy rates, if we make a certain reasonable assumption (see below).

The probabilities for singlets and doublets of spine states $P(s_{i})$ and $P(s_{i},s_{i+1})$ 
can be obtained from Eq. (1) by summations over all other states in other synapses as

$P(s_{i})= \sum_{s_{1}}... \sum_{s_{i-1}} \sum_{s_{i+1}} .... \sum_{s_{N}} P(s_{1},...,s_{i},...,s_{N})$,

and similarly

$P(s_{i},s_{i+1})= \sum_{s_{1}}... \sum_{s_{i-1}} \sum_{s_{i+2}} .... \sum_{s_{N}}
P(s_{1},...,s_{i},...,s_{N})$.

As a result we obtain two equations

\begin{eqnarray}
 \frac{d P(s_{i})}{dt}= \sum_{s_{i-1}}\sum_{s_{i+1}} \sum_{s'_{i}}
 \Big[ w_{s_{i},s'_{i}}(s_{i-1},s_{i+1}) P(s_{i-1},s'_{i},s_{i+1}) 
 - w_{s'_{i},s_{i}}(s_{i-1},s_{i+1}) P(s_{i-1},s_{i},s_{i+1}) \Big],
\end{eqnarray}\\
which is valid for $i=2,...,N-1$, and

\begin{eqnarray}
 \frac{d P(s_{i},s_{i+1})}{dt}= \sum_{s_{i-1}}\sum_{s'_{i}} 
  \left[ w_{s_{i},s'_{i}}(s_{i-1},s_{i+1}) P(s_{i-1},s'_{i},s_{i+1})
 - w_{s'_{i},s_{i}}(s_{i-1},s_{i+1}) P(s_{i-1},s_{i},s_{i+1}) \right]
  \nonumber    \\ 
+ \sum_{s_{i+2}}\sum_{s'_{i+1}} 
  \left[ w_{s_{i+1},s'_{i+1}}(s_{i},s_{i+2}) P(s_{i},s'_{i+1},s_{i+2})
 - w_{s'_{i+1},s_{i+1}}(s_{i},s_{i+2}) P(s_{i},s_{i+1},s_{i+2}) \right].
\end{eqnarray}\\
which is valid for $i=2,...,N-2$. For the boundary probabilities, i.e.,
for the dynamics of $P(s_{1}), P(s_{N})$ and $P(s_{1},s_{2}), P(s_{N-1},s_{N})$ we
have similar equations, except we drop the boundary terms $s_{0}, s_{N+1}$ as
they are nonexistent.

Equations (4) and (5)  for the dynamics of $P(s_{i})$ and  $P(s_{i},s_{i+1})$ involve
additionally the probabilities of spine triplets $P(s_{i-1},s_{i},s_{i+1})$ and
$P(s_{i},s_{i+1},s_{i+2})$, and thus they do not form a closed system of equations.
To close these equations, we use the so-called ``pair approximation'' for probabilities.
The main idea in this approximation is that the biggest influence on a given synapse is
exerted only by the nearest neighbor synapses, and the effects from remote neighbors can be
neglected, as is implied by the form of the transition rate $w_{s_{i},s'_{i}}(s_{i-1},s_{i+1})$.
Specifically, for 3 neighboring synapses indexed spatially as $i-1, i, i+1$, the dynamic
of synapse $i-1$ depends directly only on the state of synapse $i$, and the influence of
$i+1$ synapse can be neglected as coming from the remote site. In terms of probabilities,
this can be written as

\begin{eqnarray}
P(s_{i-1},s_{i},s_{i+1}) \approx 
P(s_{i-1},s_{i})P(s_{i},s_{i+1})/P(s_{i}), 
\end{eqnarray}\\
where we used the approximation for the conditional probability
$P(s_{i-1}|s_{i},s_{i+1})\approx P(s_{i-1}|s_{i})$, and the fact that
$P(s_{i-1}|s_{i})= P(s_{i-1},s_{i})/P(s_{i})$. Thus the probabilities of the spine triplets
can be effectively written as combinations of the probabilities for spine singlets and
doublets, which forms the essence of the pair approximation. A similar expression can be
obtained for synapses with other combination of indexes.

The above pair approximation allows us to write the dynamics of probabilities
$P(s_{i})$ and $P(s_{i},s_{i+1})$ as

\begin{eqnarray}
\frac{d P(s_{i})}{dt}= \sum_{s_{i-1}}\sum_{s_{i+1}} \sum_{s'_{i}}
 \Big[ w_{s_{i},s'_{i}}(s_{i-1},s_{i+1}) \frac{P(s_{i-1},s'_{i})P(s'_{i},s_{i+1})}{P(s'_{i})} 
%\nonumber  \\
  -  w_{s'_{i},s_{i}}(s_{i-1},s_{i+1})\frac{P(s_{i-1},s_{i})P(s_{i},s_{i+1})}{P(s_{i})} \Big]
\end{eqnarray}\\
for $i=2,...,N-1$, and

\begin{eqnarray}
\frac{d P(s_{i},s_{i+1})}{dt}= \sum_{s_{i-1}}\sum_{s'_{i}} 
  \left[ w_{s_{i},s'_{i}}(s_{i-1},s_{i+1}) \frac{P(s_{i-1},s'_{i})P(s'_{i},s_{i+1})}{P(s'_{i})}
 - w_{s'_{i},s_{i}}(s_{i-1},s_{i+1}) \frac{P(s_{i-1},s_{i})P(s_{i},s_{i+1})}{P(s_{i})} \right]
 \nonumber  \\   
+ \sum_{s_{i+2}}\sum_{s'_{i+1}} 
  \left[ w_{s_{i+1},s'_{i+1}}(s_{i},s_{i+2}) \frac{P(s_{i},s'_{i+1})P(s'_{i+1},s_{i+2})}{P(s'_{i+1})}
 - w_{s'_{i+1},s_{i+1}}(s_{i},s_{i+2}) \frac{P(s_{i},s_{i+1})P(s_{i+1},s_{i+2})}{P(s_{i+1})} \right],
\end{eqnarray}\\
for $i=2,...,N-2$. Similar expressions can be written for the boundary probabilities with
$i=1$ and $i=N$.

It is clear that Eqs. (7,8) for the dynamics of $P(s_{i})$ and $P(s_{i},s_{i+1})$
form the closed system of equations, since now they only depend on each other.
Importantly, the number of equations in the reduced dynamics (7,8) is only $20N-16$, which
is linear in $N$ and thus much smaller than the original $4^{N}$ equations, and hence feasible
for numerical analysis. These two types of probabilities are sufficient to compute quantities
of interest, which are associated with LTP induction, such as memory trace and its duration,
the average sizes of spines, and the rates of information gain (Kullback-Leibler divergence)
and energy dissipated (entropy production rate). However, first we check the accuracy of
the pair approximation.

\vspace{0.4cm}

\noindent{\large \bf 2.3. Validity of the pair approximation.}

In this section we check how accurate is the pair approximation by considering
a small system of dendritic spines with $N=4$, for which one can find an exact
numerical solution for the dynamics in Eq. (1). Our goal is to compare this exact
solution with its approximation given by Eqs. (7) and (8).

Numerical calculations indicate that the pair approximation is very accurate,
as exact and approximate probabilities are practically indistinguishable (Fig. 2A),
even for very strong couplings between spines ($\gamma= -0.9$ and $\gamma=0.9$).
Moreover, the pair approximation is well defined, since it preserves positivity
of all probabilities and their normalization (Fig. 2A,B). Additionally, as an
example of the main observable used in this study, we also compared entropy
production rates computed for the exact dynamics in Eq. (1), denoted as EPR$_{ex}$,
with that computed from the approximate dynamics in Eqs. (7) and (8) and
denoted as EPR$_{pa}$ (the formulas for the exact and approximate EPR are
given, respectively, in Eqs. (10-13) and Eqs. (16-19)).
Both entropy production rates are also essentially
indistinguishable, with a small difference at most $0.6 \%$ (Fig. 2C).

In order to give a measure of the pair approximation accuracy
we introduce the ratio $R$ for $N=4$ spines, defined as

\begin{eqnarray}
 R(s_{1},s_{2},s_{3},s_{4}) = 
  \frac{P_{pa}(s_{1},s_{2})P_{pa}(s_{2},s_{3})P_{pa}(s_{3},s_{4})}
        {P_{pa}(s_{2})P_{pa}(s_{3})P_{ex}(s_{1},s_{2},s_{3},s_{4})},
\end{eqnarray}\\
where the subscript $pa$ refers to the pair approximation (Eqs. 7 and 8), while
$ex$ corresponds to the exact solution (Eq. 1).
When $R$ approaches 1, then the pair approximation matches
the exact solution perfectly. The larger the deviation of $R$
from unity, the less accurate is the approximation. This follows from the
form of pair approximation for 4 spines, i.e.,
$P(s_{1},s_{2},s_{3},s_{4})\approx
P(s_{1},s_{2})P(s_{2},s_{3})P(s_{3},s_{4})/[P(s_{2})P(s_{3})]$.
To have a global numerical accuracy, we have to average $R$ over all
states, which yields a mean ratio  
 $\langle R\rangle = 
\frac{1}{4^{4}} \sum_{s_{1},...,s_{4}}  R(s_{1},...,s_{4})$,
and its standard deviation
$\mbox{SD(R)}= \sqrt{\langle R^{2}\rangle - \langle R\rangle^{2}}$,
serving as a global error.
In Fig. 3, we show that $\langle R\rangle$ is very close to 1, and
SD(R) is generally small, at most $0.1$.

Taken together, the numerical results in Figs. 2 and 3 indicate that the pair approximation
derived in this study is quite accurate, and its accuracy is preserved in time.
For an analytical example related to the pair approximation, see Appendix B.

%\vspace{0.4cm}
\newpage

{\Large \bf 3. Derivation of entropy production rate as an energy cost
for interacting spines.}  

\vspace{0.35cm}

Energy expenditure of synaptic plasticity is associated with transitions between
different states of a dendritic spine. The faster the transitions, the more energy is
used, and vice versa. Generally, a spine is in a thermodynamic nonequilibrium with its
environment, and thus the energy cost is strictly related to entropy production rate of
the spine (for general ideas of nonequilibrium thermodynamics, see: Nicolis and Prigogine,
(1977), and Peliti and Pigolotti (2021)). Specifically, we assume that the energy rate
associated with plasticity processes in dendritic spines is equal to the entropy production
associated with stochastic transitions between spine mesoscopic states,
similar as in (Karbowski 2019). In our case of $N$ dendritic
spines, the entropy production rate of the whole system $\mbox{EPR}(s_{1},...,s_{N})$
is given by a general formula for the entropy production
(Schnakenberg 1976; Maes et al 2000; Seifert 2012; Van den Broeck and Esposito 2015):

\begin{eqnarray}
\mbox{EPR}(s_{1},...,s_{N})= \sum_{i=1}^{N} \mbox{EPR}_{i}
\end{eqnarray}\\ 
where $\mbox{EPR}_{i}$ is the individual entropy productions of each interacting spine

\begin{eqnarray}
\mbox{EPR}_{i}= \frac{\epsilon}{2} \sum_{s_{1},...,s_{N}} \sum_{s'_{i}}   
  \left[ w_{s_{i},s'_{i}}(s_{i-1},s_{i+1})P(s_{1},...,s'_{i},...,s_{N})
  - w_{s'_{i},s_{i}}(s_{i-1},s_{i+1})P(s_{1},...,s_{i},...,s_{N}) \right]     \nonumber  \\
 \times \ln \frac{w_{s_{i},s'_{i}}(s_{i-1},s_{i+1})P(s_{1},...,s'_{i},...,s_{N})}
   {w_{s'_{i},s_{i}}(s_{i-1},s_{i+1})P(s_{1},...,s_{i},...,s_{N})} 
\end{eqnarray}\\ 
for $i=2,...,N-1$, and for the boundary terms

\begin{eqnarray}
\mbox{EPR}_{1}= \frac{\epsilon}{2} \sum_{s_{1},...,s_{N}} \sum_{s'_{1}}
  \left[ w_{s_{1},s'_{1}}(s_{2})P(s'_{1},s_{2},...,s_{N})
  - w_{s'_{1},s_{1}}(s_{2})P(s_{1},s_{2},...,s_{N}) \right]   \nonumber \\ 
 \times \ln \frac{w_{s_{1},s'_{1}}(s_{2})P(s'_{1},s_{2},...,s_{N})}
   {w_{s'_{1},s_{1}}(s_{2})P(s_{1},s_{2},...,s_{N})} 
\end{eqnarray}\\
and

\begin{eqnarray}
\mbox{EPR}_{N}= \frac{\epsilon}{2} \sum_{s_{1},...,s_{N}} \sum_{s'_{N}}
  \left[ w_{s_{N},s'_{N}}(s_{N-1})P(s_{1},s_{2},...,s'_{N})
  - w_{s'_{N},s_{N}}(s_{N-1})P(s_{1},s_{2},...,s_{N}) \right]   \nonumber \\ 
\times \ln \frac{w_{s_{N},s'_{N}}(s_{N-1})P(s_{1},s_{2},...,s'_{N})}
   {w_{s'_{N},s_{N}}(s_{N-1})P(s_{1},s_{2},...,s_{N})},
\end{eqnarray}\\
where $\epsilon$ is the energy scale for various biophysical processes taking place
inside a typical dendritic spine and related to molecular plasticity. Its value was estimated
at about $\epsilon\approx 4.6\cdot 10^{5}$ kT (or $2.3\cdot 10^{4}$ ATP molecules),
where $k$ is the Boltzmann constant and $T$ is the absolute brain temperature
(see, Karbowski 2021). In a nutshell, these numbers can be understood by considering
that a typical dendritic spine contains roughly $\sim 10^{4}$ proteins, each with a few
degree of freedom corresponding to the number of phosphorylation sites (Sheng and
Hoogenraad, 2007).

As before, we explore the pair approximation in Eqs. (11-13), which in the case of
$N$ spines takes the form

\begin{eqnarray}
P(s_{1},...,s_{i-1},s'_{i},s_{i+1},...,s_{N}) \approx
\frac{P(s_{1},s_{2})...P(s_{i-1},s_{i}')P(s_{i}',s_{i+1})...P(s_{N-1},s_{N})}
  {P(s_{2})...P(s_{i-1})P(s_{i}')P(s_{i+1})...P(s_{N-1})}.
\end{eqnarray}\\
This is a straightforward generalization of formula (6), which can be easily
verified. Application of Eq. (14) leads to simplification of the ratio of
probabilities under the logarithm in Eq. (11) as

\begin{eqnarray}
  \frac{P(s_{1},...,s_{i-1},s'_{i},s_{i+1},...,s_{N})}
{P(s_{1},...,s_{i-1},s_{i},s_{i+1},...,s_{N})}
  \approx
\frac{P(s_{i-1},s_{i}')P(s_{i}',s_{i+1})P(s_{i})}
{P(s_{i-1},s_{i})P(s_{i},s_{i+1})P(s_{i}')},
\end{eqnarray}\\
which allows us to perform summation over almost all
states $s_{1}, s_{2}, ... , s_{N}$ except the few in Eqs. (11-13).
This step produces the final expression for the approximate total entropy
production rate $\mbox{EPR}(s_{1},...,s_{N})$ of our interacting spines:

\begin{eqnarray}
\mbox{EPR}(s_{1},...,s_{N})= \sum_{i=1}^{N} \mbox{EPR}_{i}
\end{eqnarray}\\ 
where

\begin{eqnarray}
\mbox{EPR}_{i}\approx \frac{\epsilon}{2} \sum_{s_{i-1},s_{i+1}} \sum_{s_{i},s'_{i}}
\left[ w_{s_{i},s'_{i}}(s_{i-1},s_{i+1})\frac{P(s_{i-1},s'_{i})P(s'_{i},s_{i+1})}
  {P(s'_{i})} - w_{s'_{i},s_{i}}(s_{i-1},s_{i+1})\frac{P(s_{i-1},s_{i})P(s_{i},s_{i+1})}
  {P(s_{i})} \right]   \nonumber \\ 
 \times \ln \frac{w_{s_{i},s'_{i}}(s_{i-1},s_{i+1})P(s_{i-1},s'_{i})P(s'_{i},s_{i+1})P(s_{i})}
   {w_{s'_{i},s_{i}}(s_{i-1},s_{i+1})P(s_{i-1},s_{i})P(s_{i},s_{i+1})P(s'_{i})} 
\end{eqnarray}\\ 
for $i=2,...,N-1$, and for the boundary terms

\begin{eqnarray}
\mbox{EPR}_{1}\approx \frac{\epsilon}{2} \sum_{s_{1},s_{2}} \sum_{s'_{1}}
  \left[ w_{s_{1},s'_{1}}(s_{2})P(s'_{1},s_{2})
  - w_{s'_{1},s_{1}}(s_{2})P(s_{1},s_{2}) \right]  \nonumber  \\ 
 \times \ln \frac{w_{s_{1},s'_{1}}(s_{2})P(s'_{1},s_{2})}
   {w_{s'_{1},s_{1}}(s_{2})P(s_{1},s_{2})} 
\end{eqnarray}\\
and

\begin{eqnarray}
\mbox{EPR}_{N}\approx \frac{\epsilon}{2} \sum_{s_{N-1},s_{N}} \sum_{s'_{N}}
  \left[ w_{s_{N},s'_{N}}(s_{N-1})P(s_{N-1},s'_{N})
  - w_{s'_{N},s_{N}}(s_{N-1})P(s_{N-1},s_{N}) \right]   \nonumber \\ 
 \times \ln \frac{w_{s_{N},s'_{N}}(s_{N-1})P(s_{N-1},s'_{N})}
   {w_{s'_{N},s_{N}}(s_{N-1})P(s_{N-1},s_{N})}.
\end{eqnarray}\\
Note that the total entropy production of all spines $\mbox{EPR}(s_{1},...,s_{N})$
is determined exclusively in terms of the two types of probabilities considered in
Eqs. (7) and (8), i.e., one- and two-point probabilities.
It is also interesting to mention that the form of the approximated EPR in Eq. (17)
can be also deduced instantly from the form of the approximated dynamics for
probabilities in Eq. (7). This is possible if one realizes that the expression
in the bracket on the right in Eq. (7) represents a probability flux.

The total energy $E$ used by all spines for LTP induction and its maintenance up
to recovery (during synaptic stimulation and post stimulation) is the energy needed
to keep the memory trace above the threshold. $E$ is the total energy cost of LTP,
and it is defined as

\begin{eqnarray}
E= \int_{0}^{\overline{T}_{M}} dt \; \mbox{EPR}(t),
\end{eqnarray}\\
where $t=0$ relates to the onset of stimulation, $\mbox{EPR}(t)$ is the total entropy
production rate given by Eqs. (16-19), and $\overline{T}_{M}= T_{M}+\tau_{1}+\tau_{2}$,
where $T_{M}$ is the memory time, and $\tau_{1}+\tau_{2}$ is the duration of the
stimulation (learning phase; see Eq. (3)). The energy used solely for LTP induction
and maintenance is denoted as $E_{ltp}$, and it is given by
$E_{ltp}= E - \mbox{EPR}_{0}\overline{T}_{M}$, where $\mbox{EPR}_{0}$ is the baseline
entropy production rate of all spines.

\newpage
%\vspace{0.4cm}

{\Large \bf 4. Definition of memory trace and memory time.}

We define the signal associated with dendritic spine activation as

\begin{eqnarray}
S= \frac{1}{N} \sum_{i=1}^{N} s_{i}.
\end{eqnarray}\\
The signal at a steady state (baseline) is denoted as $S_{ss}$.
Memory trace $MT$ is defined as the average normalized signal to noise ratio.
The normalized signal is simply its deviation from the steady state or baseline.
Consequently, the memory trace takes the form:

\begin{eqnarray}
  MT= \frac{(\langle S\rangle - \langle S_{ss}\rangle)}
  {\sqrt{\langle S^{2}\rangle - \langle S\rangle^{2}}} ,
\end{eqnarray}\\
where $\langle S^{n}\rangle= \sum_{s_{1}}...\sum_{s_{N}} S^{n} P(s_{1},...,s_{N})$
for $n=1,2$. The explicit forms for the signals and variance of the signal are

\begin{eqnarray}
\langle S\rangle = \frac{1}{N} \sum_{i=1}^{N} \langle s_{i}\rangle,
\end{eqnarray}\\
\begin{eqnarray}
\langle S\rangle_{ss} = \frac{1}{N} \sum_{i=1}^{N} \langle s_{i}\rangle_{ss},
\end{eqnarray}\\
\begin{eqnarray}
  \langle S^{2}\rangle - \langle S\rangle^{2} =
  \frac{1}{N^{2}} \sum_{i=1}^{N} \big[\langle s_{i}^{2}\rangle - \langle s_{i}\rangle^{2}\big] 
  + \frac{1}{N^{2}} \sum_{i\ne j}\sum_{j=1}^{N} \big[ \langle s_{i}s_{j}\rangle -
  \langle s_{i}\rangle\langle s_{j}\rangle \big], 
\end{eqnarray}\\
where $\langle s_{i}^{n}\rangle= \sum_{s_{i}} s^{n}_{i} P(s_{i})$,
$\langle s_{i}\rangle_{ss}= \sum_{s_{i}} s_{i} P_{ss}(s_{i})$, and
$\langle s_{i}s_{j}\rangle= \sum_{s_{i}}\sum_{s_{j}}  s_{i}s_{j} P(s_{i},s_{j})$,
where $P_{ss}(s_{i})$ is the probability distribution at baseline state.
The first sum on the right in Eq. (25) is the sum of variances of individual spines,
while the second sum is the total cross correlation of spines. 
In the pair approximation for the probability, the last sum associated with the
correlations simplifies, as only the cross correlations between neighboring spines
provide nonzero contributions, since generally 
$P(s_{i},s_{j}) \approx P(s_{i})P(s_{j})$ for $|i-j| \ge 2$.

Memory time $T_{M}$ is defined as the time $t$ after stimulation for which memory trace
$MT$ is in a declining phase and assumes value 1 (Fusi et al 2005; Leibold and Kempter 2008;
Karbowski 2019). This is the moment in time when normalized signal becomes comparable to
its noise component.

%\vspace{0.5cm}
\newpage

{\Large \bf 5. Derivation of information gain for interacting spines.}  

\vspace{0.35cm}

Information gain $I$, for all spines, right after the end of LTP (i.e., when memory
trace decays to the noise level $MT= 1$) is defined as Kullback-Leibler divergence
at time $t=\overline{T}_{M}$, i.e. $\mbox{KL}(t=\overline{T}_{M})$, between the baseline
steady-state initial probability $P(s_{1},...,s_{N})_{ss}$ at time $t=0$
(before LTP stimulation) and final probability at time $t= \overline{T}_{M}$,
i.e., $P(s_{1},...,s_{N};t=\overline{T}_{M})$. Its form is given by

\begin{eqnarray}
I\equiv \mbox{KL}\left( P(s_{1}, ...,s_{N};t=\overline{T}_{M}) || P(s_{1}, ..., s_{N})_{ss} \right)
\nonumber  \\ 
= \sum_{s_{1},...,s_{N}} P(s_{1},...,s_{N};t=\overline{T}_{M})
  \ln \frac{ P(s_{1},...,s_{N};t=\overline{T}_{M}) }{ P(s_{1},...,s_{N})_{ss} }.
\end{eqnarray}\\

For the pair approximation, i.e., applying the approximation (14) for the
probabilities, information gain $I$ takes the form

\begin{eqnarray}
 I =  \sum_{i=1}^{N-2} \sum_{s_{i}} \sum_{s_{i+1}} P(s_{i},s_{i+1})
  \ln \frac{P(s_{i},s_{i+1})P_{ss}(s_{i+1})}{P_{ss}(s_{i},s_{i+1})P(s_{i+1})}  \nonumber  \\ 
  +  \sum_{s_{N-1}} \sum_{s_{N}} P(s_{N-1},s_{N})
  \ln \frac{P(s_{N-1},s_{N})}{P_{ss}(s_{N-1},s_{N})}, 
\end{eqnarray}\\
which is used in the computations.

The rate of information gain is equivalent to the rate of Kullback-Leibler KL divergence
at arbitrary time $t$, i.e., $\mbox{KL}(t)$ (given by Eq. (26) but for arbitrary $t$).
We are interested in the rate of information gain, since we want to compare it
directly to the entropy production rate, which has a similar information-theoretic meaning.
The rate of KL, which we denote as $\mbox{KLR}$, is given by $\mbox{KLR}= d\mbox{KL}(t)/dt$.
We find

\begin{eqnarray}
\mbox{KLR} = \sum_{s_{1},...,s_{N}} \frac{d P(s_{1},...,s_{N})}{dt}
\ln \frac{P(s_{1},...,s_{N})}{P_{ss}(s_{1},...,s_{N})}, 
\end{eqnarray}\\
where we used the fact that   $\sum_{s_{1},...,s_{N}} d P(s_{1},...,s_{N})/dt = 0$.
The next step is to substitute Eq. (1) for $\frac{d P(s_{1},...,s_{N})}{dt}$,
and perform summations over almost all states $s_{1},..., s_{N}$, except the
few, similarly to the calculation for EPR. Finally, we use the pair approximation.
As a result we obtain the total rate of information gain for all spines as

\begin{eqnarray}
 \mbox{KLR} =  \sum_{s_{1},s'_{1}} \sum_{s_{2}} 
 \left[ w_{s_{1},s'_{1}}(s_{2})P(s'_{1},s_{2}) - w_{s'_{1},s_{1}}(s_{2})P(s_{1},s_{2}) \right]   
  \ln \frac{P(s_{1},s_{2})}{P_{ss}(s_{1},s_{2})}      \nonumber  \\ 
+ \sum_{s_{N},s'_{N}} \sum_{s_{N-1}} 
 \left[ w_{s_{N},s'_{N}}(s_{N-1})P(s_{N-1},s_{N}') - w_{s'_{N},s_{N}}(s_{N-1})P(s_{N-1},s_{N}) \right]   
  \ln \frac{P(s_{N-1},s_{N})}{P_{ss}(s_{N-1},s_{N})}    \nonumber   \\ 
 + \sum_{i=2}^{N-1} \sum_{s_{i},s'_{i}} \sum_{s_{i-1},s_{i+1}} 
\left[ w_{s_{i},s'_{i}}(s_{i-1},s_{i+1})\frac{P(s_{i-1},s'_{i})P(s'_{i},s_{i+1})}
  {P(s'_{i})} - w_{s'_{i},s_{i}}(s_{i-1},s_{i+1})\frac{P(s_{i-1},s_{i})P(s_{i},s_{i+1})}
  {P(s_{i})} \right]   \nonumber  \\ 
  \times \ln \frac{P(s_{i-1},s_{i})P(s_{i},s_{i+1})P_{ss}(s_{i})}
   {P_{ss}(s_{i-1},s_{i})P_{ss}(s_{i},s_{i+1})P(s_{i})} 
\end{eqnarray}\\ 
As can be seen, the rate of information gain KLR depends, similar to EPR, on the transition
rates between states in all spines. In all figures, we plot KLR per spine, i.e., KLR/$N$.
Finally, note that the information gain $I$ is the temporal integral of KLR from $t=0$
to $t= \overline{T}_{M}$, i.e.,  $I= \int_{0}^{\overline{T}_{M}} dt \; \mbox{KLR}(t)$.

%\vspace{0.5cm}
\newpage

{\Large \bf 6. Numerical results on memory trace, information gain, and energy cost
 during LTP.}  

\vspace{0.4cm}

We divide the global dynamics of our system of interacting dendritic spines into two
stages. The first stage relates to approaching and reaching a steady-state, starting
from random initial conditions (with $f(t)=0$ for each spine).
This steady-state is a thermodynamic non-equilibrium
steady-state that uses some small but nonzero energy (small but nonzero EPR), and can
be thought as the state in which some background information is written in synapses.
After reaching the steady-state, we start a second stage associated with spine stimulation,
which we call LTP (long-term potentiation) phase. This stage consists of a brief stimulation
of the synaptic system by amplifying the transition rates by the function $f(t)$ present
in Eqs. (2) and (3), and then observation of the system recovery to the steady state,
with simultaneous recording of the most important observables. Stimulation of synapses
is done by turning on the amplifier function $f(t)$, which amplifies the transition
rates between synaptic states (Eqs. 2 and 3). We call this stimulation phase as ``learning''
phase, and recovery phase as ``memory'' phase.

\vspace{0.4cm}

\noindent{\large \bf 6.1. Dynamics of memory trace, information and energy rates, and
  synaptic size associated with LTP induction.}

Following LTP induction (starting at time $t=0$ in $f(t)$ function) memory trace $MT$
contained in synapses behaves differently than the rates of information gain (Kullback-Leibler
rate KLR) and energy (entropy production rate EPR) (Fig. 4). Initially all three quantities
increase sharply, similarly as $f(t)$, but later their dynamics diverge. Specifically,
memory trace exhibits a long temporal tail, i.e., it decays much slower than the stimulation
function $f(t)$, with longer tail for positive cooperation ($\gamma > 0$) between neighboring
spines than for negative cooperation (Fig. 4B). On the other hand, both KLR and EPR (per synapse)
decay extremely fast to their baseline values, much faster then $f(t)$ (Fig. 4C,D). This strongly
suggests that keeping memory trace high does not require large rates of energy (EPR).
Moreover, the dynamics of KLR and EPR are very similar in shape, and their
ratio is positive only initially, when synaptic stimulation increases in time (Fig. 4E).
This behavior indicates that information amount written at synapses increases sharply only
in the beginning of LTP (learning); at later stages (memory) it weakly decreases (note
negative values of KLR/EPR). The rates of information gain (KLR) and energy (EPR) depend
weakly, and in the opposite way, on the sign of cooperativity $\gamma$; the peak of KLR is
greater for negative $\gamma$, while the peak of EPR is greater for positive $\gamma$
(Fig. 4C,D). At its peak, i.e. during stimulation, energy is consumed at rate $4.6\cdot 10^{5}$
kT/min, and information is gained at rate 0.1 bits/min, which indicates that acquiring 1 bit
at that moment in time is very expensive and costs about $4.6\cdot 10^{6}$ kT.

The dynamics of spine sizes following LTP induction are similar to the behavior of memory
trace, except that sizes stabilize at some finite level (Fig. 5). Positive cooperativity among
synapses ($\gamma > 0$) generally indicates positive correlations between them, and vice versa
(Fig. 5C). Consequently, positive correlations lead to higher mean spine sizes than negative
correlations (Fig. 5B). An opposite effect is seen for the ratio of KLR and spine size;
higher peaks are observed for negative cooperativity between synapses (Fig. 5D). Interestingly,
normalized correlations are more variable during learning and memory phases for negative
cooperativity than for positive one (Fig. 5C).

\vspace{0.4cm}

\noindent{\large \bf 6.2. Memory time, spine size, information gain after LTP, and their
 energy costs as functions of synaptic cooperativity.}

Memory time $T_{M}$ grows monotonically but very weakly with cooperativity $\gamma$ up
to a point where $\gamma$ is close to its maximal value 1 (Fig. 6A). In that regime of very
high positive cooperativity, $T_{M}$ increases sharply with $\gamma$. The opposite dependence
on $\gamma$ is present for information gain $I$ and average spine size (Fig 6B, E).
$I$ generally decreases with $\gamma$ (for negative $\gamma$ the decay is stronger than
for positive $\gamma$; Fig. 6B), whereas mean spine size monotonically increases with
$\gamma$ (increase is stronger for $\gamma$ close to 1; Fig. 6E). Total energy $E$ consumed
during LTP and its part $E_{ltp}$ related solely to LTP depend nonmonotonically on
cooperativity $\gamma$, exhibiting broad maxima for $\gamma= 0$ (Fig. 6C,D). For $\gamma$
close to its maximal value 1, $E$ and $E_{ltp}$ behave in opposite ways: the former increases
while the latter decreases with $\gamma$. This suggests that the cost of LTP alone drops
for very high cooperativity.

How do these results translate to energy and structural efficiency of memory lifetime and
information gain? Figure 7 provides the answers. The ratios of memory time and the energies,
i.e., $T_{M}/E$ and $T_{M}/E_{ltp}$, are essentially constant for $\gamma$ up to $\sim 0.8$
(Fig. 7A), suggesting that memory time and both of these energies grow proportionally with
cooperativity for a wide range of $\gamma$. For larger $\gamma$ these ratios grow significantly
with $\gamma$, especially $T_{M}/E_{ltp}$, indicating that energy efficiency of memory time is
enhanced in the regime of very high positive synaptic cooperativity (Fig. 7A).

Energy efficiency of information gain is more complex (Fig. 7B). Generally, the ratio of
information to total energy during LTP $I/E$ decreases monotonically with $\gamma$,
meaning that efficiency of $I$ is maximal for negative cooperativity between synapses.
On the other hand, the ratio of information to energy solely to LTP, i.e.,  $I/E_{ltp}$
as a function of $\gamma$ has a U-shape, with large values for both high negative and high
positive cooperativity. The latter means that information efficiency in that energy currency
has two regimes of higher values (Fig. 7B). However, it should be emphasized that the overall
energy efficiency of information gain is rather low, at $(5-10)\cdot 10^{-4}$ bits/$\epsilon$
or $(1-2)\cdot 10^{-9}$ bits/kT, i.e., 1 bit of stored memory (after decline of LTP, i.e. after
time $\sim T_{M}$) costs about $10^{9}$ kT (see also below).

Structural efficiency (or energy efficiency of transmission) of memory time and information
gain exhibit different behavior as functions of synaptic cooperativity (Fig. 7 C,D). The
ratio of information gain to mean spine size $I/\langle S\rangle$ decreases monotonically
with $\gamma$, which indicates that structural efficiency of information is the largest
for negative cooperativity (Fig. 7D), similar to the (total) energy efficiency of $I$
(Fig. 7B). The ratio of memory time to mean spine size $T_{M}/\langle S\rangle$ decreases
slightly with increasing cooperativity from negative values of $\gamma$, but for
$\gamma \approx 0.4$ this ratio increases sharply with $\gamma$ (Fig. 7C).
This result shows that structural efficiency of memory time is the highest for strong
positive synaptic cooperativity (Fig. 7C), where spine normalized correlations reach
values $0.2-0.3$ (Fig. 7E).

\vspace{0.4cm}

\noindent{\large \bf 6.3. Sparse representations of synaptic memory and information
  are more energy efficient.}

Next we investigate how memory time, information gain, average spine size, and energy
cost depend on the fraction $p_{act}$ of stimulated synapses by presynaptic neurons (Fig. 8).
Memory time $T_{M}$, energy cost solely due to LTP $E_{ltp}$, and mean spine size grow
monotonically with $p_{act}$, though the first one saturates for larger $p_{act}$ (Fig 8 A,C,D).
In contrast, information gain $I$ and total energy cost $E$ display non-monotonic behavior:
the former has a maximum, while the latter a minimum for a small fraction of active synapses
(Fig. 8 B,C). In terms of energy efficiency, the ratios of $T_{M}/E, T_{M}/E_{ltp}$ and
$I/E, I/E_{ltp}$ have maxima at around the same small fraction $p_{act}$, regardless of the
sign of synaptic cooperativity $\gamma$ (Fig. 9A,B). This means that there exist an optimal
percentage of activated synapses on a dendrite that yields the highest information gain and
memory lifetime per invested energy (whether total or only due to LTP). For that optimal
$p_{act}$, the energy cost of 1 bit of stored information is about $10^{7}$ kT, which is
much lower (and thus more efficient) than for values $p_{act}$ away from the optimality.
Interestingly, the normalized correlations between spines are essentially independent
of $p_{act}$ (Fig. 9E).

Structural efficiency of information gain and memory time is qualitatively similar to
their energy efficiency (Fig. 9). The information per spine size $I/\langle S\rangle$
has a similar sharp peak as $I/E$ for low $p_{act}$ (Fig. 9D). However, memory time per
spine size $T_{M}/\langle S\rangle$. has a much broader maximum at higher values of
$p_{act}$ (Fig. 9C).

Taken together, these results indicate that energetic and structural efficiency of
information and its duration in synapses can be optimized for low fractions of activated
synapses during LTP. In other words, acquiring and storing of synaptic information can be
most efficient by using sparse synaptic representations, regardless of the nature of
synaptic cooperativity.

%\vspace{2.8cm}

\vspace{0.4cm}

\noindent{\large \bf 6.4. Memory time, spine size, information gain, and their
 energy costs as functions of strength and duration of stimulation.}

In Figs. 10 and 11, we show how memory time, information gain, energy cost, and mean spine
size depend on the duration of stimulation $\tau_{1}$ (decay time of the stimulation).
Memory time $T_{M}$ and its energy costs $E, E_{ltp}$ both grow proportionally with $\tau_{1}$
(Fig. 10A,C), such that their ratios $T_{M}/E$ and $T_{M}/E_{ltp}$ are almost constant,
although with a weak increasing trend (Fig. 11A). Information gain $I$ as well as its energy
efficiencies $I/E, I/E_{ltp}$ decrease with $\tau_{1}$ for small $\tau_{1}$, but for larger
$\tau_{1}$ all of these quantities saturate at some small level (Figs. 10B and 11B). Average
spine size shows a similar saturation effect after a small initial increase (Fig. 10D).
In terms of structural efficiency, $T_{M}/\langle S\rangle$ grows linearly with $\tau_{1}$
(Fig. 11C), and $I/\langle S\rangle$ first decreases with $\tau_{1}$ and then saturates
(Fig. 11D). The results in Fig. 11 indicate that longer stimulation times are not particularly
beneficial for energy efficiency of memory lifetime and information gain (Fig. 11A,B). However,
longer stimulation may be advantageous for structural efficiency of memory duration (Fig. 11C),
though not for information $I$ (Fig. 11D).

In Figs. 12 and 13, we present the dependence of memory time, information gain,
energy cost, and mean spine size on the amplitude of stimulation $A$. Memory time $T_{M}$
and energy costs both grow weakly but saturate with $A$ (Fig. 12 A,C). On the other hand,
$I$ and spine size $\langle S\rangle$ stay almost constant (Fig. 12 B,D). These results
translate into very weak variability of energy and structural efficiencies of memory time and
information gain on $A$, which are close to constancy, except the ratio $T_{M}/\langle S\rangle$
that exhibits an increasing trend, but with a saturation (Fig. 13). This suggests that too
strong stimulations are also not advantageous over weaker stimulations for efficiency of
information gain and its duration.

\vspace{0.4cm}

\noindent{\large \bf 6.5. Influence of synaptic number on the efficiencies of memory
  lifetime and information gain.}

In Fig. 14, we show that energy efficiency of both memory lifetime $T_{M}$ and information
gain $I$ exhibit a decreasing trend with increasing the number of dendritic spines $N$.
The biggest drop in efficiency is for changing $N$ from 10 to $\sim 2000$ (Fig. 14A,B).
For higher values of $N$, the rate of decline is much slower. This result suggests that having
large number of spines on a dendrite is generally highly inefficient in terms of energy for
storing information. For example, for $N=10$ we obtain $T_{M}/E\approx 1$ min/$\epsilon$,
and $I/E\approx 4\cdot 10^{-2}$ bits/$\epsilon$, i.e., storing 1 min of memory in the spines
costs about $\epsilon\approx 4.6\cdot 10^{5}$ kT of energy, and storing 1 bit after the
degradation of LTP costs $25\epsilon\approx 10^{7}$ kT. In contrast, for $N=(8-10)\cdot 10^{3}$,
we have $T_{M}/E\approx 2\cdot 10^{-3}$ min/$\epsilon$ and $I/E\approx 10^{-4}$ bits/$\epsilon$,
which means that in this case storing 1 min of memory costs
$\sim 500\epsilon\approx 2\cdot 10^{8}$ kT, and storing 1 bit in all $N$ spines after LTP
degradation costs $10^{4}\epsilon\approx 4.6\cdot 10^{9}$ kT.

Since memory lifetime and information gain are collective variables, to determine
structural efficiencies of these variable as functions of $N$, we have to divide
$T_{M}$ and $I$ by the whole structural cost of all spines, i.e., by $N\cdot\langle S\rangle$.
Thus, the structural efficiencies in this case are the ratios $T_{M}/(N\cdot\langle S\rangle)$
and $I/(N\cdot\langle S\rangle)$ (Fig. 14 C,D), and they drop significantly with increasing $N$,
in a similar manner as the corresponding energy efficiencies (in Fig. 14 A,B).

Taken together the two types of the efficiency, it is clear that too many synapses on
a dendrite is not beneficial for information storage and its duration in terms energy
and biochemical resources.

\newpage
%\vspace{2.3cm}

{\Large \bf 7. Summary and Discussion.}

\vspace{0.3cm}

We formulated the probabilistic approach to global dynamics of interacting synapses
(on a single dendrite) and to their nonequilibrium thermodynamics to study information
processing in synaptic internal degrees of freedom.
In order to make the high dimensional system of interacting synapses computationally
tractable, we introduced the so-called pair approximation, which effectively reduces
the dimensionality and number of equations describing the system dynamics in a closed
form by considering only the probabilities of singlets and doublets of dendritic spines.  
We verified on a simplified example that this approximation provides a very good accuracy
of the exact dynamics, as well as entropy production rate. We also gave the analytical
condition for the applicability of the pair approximation in the equilibrium Ising
model in Appendix B (for a more general discussion about the accuracy of the pair
approximation, see Matsuda et al (1992), and van Baalen (2000)).

The Master equation approach combined with stochastic thermodynamics allows us to treat
information contained in synaptic states on equal footing with its energy cost by relating
them to, respectively, the rates of Kullback-Leibler divergence and entropy production.
Both of these quantities depend on state probabilities as well as on
transitions between the states, and we provided explicit formulas for their calculations. The
formalism makes it clear that every plastic transition in synaptic system is associated both
with some information processing (flow) and with some energy expenditure (entropy
production). Even stationary states out of thermodynamic equilibrium (baseline states),
in order to maintain them, require some energy (usually small), which essentially means that
keeping information always costs some energy for nonequilibrium systems even in stationary
conditions. Physically, this energy dissipation (entropy production) in stationary state is
a consequence of breaking the so-called detailed balance in probability flows, and it is
a generic feature of systems out of thermodynamic equilibrium (in the context of physics,
see: Maes et al 2000; Seifert 2012; for a dendritic spine, see: Karbowski 2019).

Our main results are: (i) Learning phase of a signal (spine stimulations) involves high
levels of both information and energy rates, which are much larger than their values
during a memory phase (Fig. 4). This indicates that keeping information (memory) is relatively
cheap in comparison to acquiring it (learning). This result on a level of many interacting
synapses is qualitatively similar to the result in (Karbowski 2019), where it was shown that
memory trace resulting from molecular transitions in a single synapse (protein phosphorylation)
decouples from energy rate after stimulation phase, leading to a relatively cheap long memory
trace of protein configurations. In our model, the maximal energy rate of spine plasticity
taking place during stimulation is $\sim 4.6\cdot 10^{5}$ kT/min (for $p_{act}=0.3$).
(ii) Memory lifetime and its energy efficiency can significantly increase their values for
very strong positive synaptic cooperativity, while the opposite is observed for information
gain right after LTP drops to its noise level (Fig. 7). This result suggests that strong local
positive correlations between neighboring spines can be beneficial for memory storage,
especially in the range $\sim 0.3$ (Fig. 7E), i.e., for values reported experimentally
(Makino and Malinow 2011). This conclusion supports the so-called ``synaptic clustering hypothesis'',
which was proposed as a mechanism for producing synaptic memory (Govindarajan et al 2006)
and enhancing its capacity (Poirazi and Mel 2001; Kastellakis and Poirazi 2019). 
(iii) There exists an optimal fraction of stimulated synapses during LTP for which energy
efficiency of both memory lifetime and information gain exhibit maxima (Fig. 9). This means
that sparse representations of learning and memory are much better in terms of energy efficiency,
and thus might be preferable by actual synapses. This is also true for structural efficiency
of information gain (Fig. 9). (iv) Energy and structural efficiencies of memory lifetime and
information gain after degradation of LTP both drop dramatically with increasing the number of
spines (Fig. 14). For example, storing 1 bit after LTP is over costs ``only'' $\sim 10^{7}$ kT
for 10 spines, and at least two orders of magnitude more, i.e., $\sim 4.6\cdot 10^{9}$ kT
for $\sim 10^{4}$ spines. Interestingly, the latter figure for energy cost is about
$\sim 10^{4}$ times larger than the cost of transmitting 1 bit through a chemical synapse
(Laughlin et al 1998). This high inefficiency of $T_{M}$ and $I$ is surprising and can
induce a question: why do neurons have so many synapses? Perhaps, the answer is that neurons
need many synapses for reliable transfer of an electric signal (short-term information)
between themselves, not necessarily for efficient storing of long-term information.

Our model is based on the assumption that a dendritic spine can be treated as the system
with discrete states, which is compatible with some morphological observations (Bourne
and Harris 2008; Montgomery and Madison 2004; Bokota et al 2016; Urban et al 2020). 
In this respect, it is similar in architecture to some previous discrete models of synapses
or dendritic spines (Fusi et al 2005; Leibold and Kempter 2008; Barrett et al 2009;
Benna and Fusi 2016). However, these models treat synaptic states quite arbitrary and
abstract, and consider mostly unidirectional transitions between the states, which makes
these models thermodynamically inconsistent (e.g. entropy production rate, equivalent to
plasticity energy rate, is ill-defined and yields infinities for unidirectional transitions).
In contrast, our model takes as a basis well defined morphological synaptic states, with
bidirectional transitions between them that are estimated based on empirical data
(Bokota et al 2016; Basu et al 2018; Urban et al 2020). The latter feature, i.e.,
bidirectional transitions, makes our model thermodynamically consistent (with finite
entropy production), as was explained in a previous model of metabolic molecular
activity in a single spine within the framework of cascade models of learning and
memory (Karbowski 2019).

In this paper, we consider two types of costs. The first is energy cost (related to entropy
production) associated with stochastic ``plastic'' transitions between different synaptic
states. The second is structural cost, defined here as proportional to average spine size and
related to biochemical cost of building a synapse. This structural cost is also proportional
to the rate of electric energy of synaptic transmission (Attwell and Laughlin 2001;
Karbowski 2009, 2012), as spine size is proportional to synaptic electric conductance or,
more commonly, synaptic weight (Kasai et al 2003).
For standard cortical conditions, i.e. for low firing rates
$\sim 1$ Hz, the energy cost of synaptic transmission is much larger than the energy cost
related to plasticity processes inside spine (Karbowski 2019). However, these two costs can
become comparable for very large firing rates, $\sim 100$ Hz (Karbowski 2021). 
There is some confusion in the literature about these two types of energy costs, and some
researchers associate the structural cost related to synapse size/weight with ``plasticity
energy cost'', by assuming the larger synaptic weight leads the higher plastic energy cost
(e.g. Li and van Rossum 2020). However, this does not have to be so, and we should make
a distinction between the two energy costs. Bigger synapses do not necessarily require larger
amounts of plasticity related energy than smaller synapses, because the transitions in bigger
synapses could be generally much slower than in smaller synapses (as is in fact reported in
some experiments; Kasai et al 2003). Thus, what mostly matter for the plasticity energy rate
are the rates of transition between internal synaptic states. On the other hand, synaptic
size/weight is always a good indicator of electric energy rate related to fast synaptic
transmission (Attwell and Laughlin 2001; Karbowski 2009, 2012).

The present model can be extended in several ways, e.g., by introducing heterogeneity
in spine interactions. That is, by allowing random signs of the cooperativity parameter $\gamma$.
However, we suspect that such modifications would not alter the general qualitative conclusions.
Finally, our model considers only the early phase of LTP, the so-called e-LTP, which generally
lasts up to a few hours and does not involve protein synthesis inside spines. Inclusion of
protein synthesis, associated with the process of memory consolidation and thus late phase of
LTP (so-called l-LTP), would require some modifications in our present model. The most important
of which is inclusion of additional variables in the probabilities characterizing spine states,
which are related to internal degrees of freedom (proteins, actin, etc). This clearly would make 
the model much more complex, and thus it remains a major challenge at present. However, we
hope that our approach of stochastic thermodynamics provides some insight about attempting to
model the interplay of information and energy during the late phase of LTP and memory
consolidation for interacting spines.

\newpage
%\vspace{2.3cm}

\vspace{0.3cm}

\noindent{\bf \large Appendix A: Stochastic model of morphological states in dendritic spines.}

\vspace{0.35cm}

Our data on dendritic spines come from cultures of rat hippocampus (Bokota et al 2016;
Basu et al 2018). We assume that each dendritic spine can be in four different morphological
states: nonexistent (lack of spine), stubby, filopodia, and mushroom. These four states
constitute the minimal number of states that can be classified and quantified on
a mesoscopic level (Bokota et al 2016; Basu et al 2018; Urban et al 2020).
Each state has a typical size, which can be characterized by several geometric parameters.
We focus on one particular parameter, spine head surface area, as an indicator of both spine
structure and function. Spine head surface area is proportional to synaptic weight (as measured
by the number of AMPA receptors; Kasai et al 2003), which relates to spine neurophysiological
function (synaptic transmission and information storage in molecular structure). On the other
hand, spine area is also a measure of its structural and metabolic costs (larger area larger
both costs). To estimate spine areas in each of the three states (for nonexistent state the size
is 0), we used the data on minimal and maximal spine head diameters from (Bokota et al 2016;
Urban et al 2020), which gave us the following numbers: $d(0)= 0$ for nonexistent,
$d(1)=0.496$ $\mu$m$^{2}$ for stubby, $d(2)=0.786$ $\mu$m$^{2}$ for filopodia/thin, and
$d(3)=1.045$ $\mu$m$^{2}$ for mushroom. 
Values of these intrinsic transition matrix are given in Table 1.
The values of global parameters are presented in Table 2.

We assume that global spine dynamics can be described as Markov chain model, in which there
are stochastic transitions between spine internal states. The general model of this kind
is given by the following master equation (e.g. Glauber 1963):

\begin{eqnarray}
  \frac{d P(s_{1},...,s_{N})}{dt}= \sum_{s'_{1},...,s'_{N}}
  W(s_{1},...,s_{N}|s'_{1},...,s'_{N}) P(s'_{1},...,s'_{N})         \nonumber  \\  
  - P(s_{1},...,s_{N}) \sum_{s'_{1},...,s'_{N}}
    W(s'_{1},...,s'_{N}|s_{1},...,s_{N})
\end{eqnarray}\\ 
where  $W(s_{1},...,s_{N}|s'_{1},...,s'_{N})$ is the multidimensional transition matrix
of the whole system of $N$ dendritic spines. We assume that transitions between the states take
place only in one of the spines at any given time unit (the rest of states in other spines do
not change in that brief time step). This means that the multidimensional
matrix $W(s_{1},...,s_{N}|s'_{1},...,s'_{N})$ can be decomposed as

\begin{eqnarray}
  W(s_{1},...,s_{N}|s'_{1},...,s'_{N})= w_{s_{1},s'_{1}}(s_{2})\delta_{s_{2}s'_{2}}...\delta_{s_{N}s'_{N}}
  + ...       \nonumber  \\  
  + w_{s_{i},s'_{i}}(s_{i-1},s_{i+1})\delta_{s_{1}s'_{1}}...\delta_{s_{i-1}s'_{i-1}}\delta_{s_{i+1}s'_{i+1}}
  ...\delta_{s_{N}s'_{N}}
  + ...       \nonumber  \\
  + w_{s_{N},s'_{N}}(s_{N-1})\delta_{s_{1}s'_{1}}...\delta_{s_{N-1}s'_{N-1}},
\end{eqnarray}\\ 
where  $w_{s_{i},s'_{i}}(s_{i-1},s_{i+1})$ are the transitions rates at individual spines, which
are dependent on the neighboring spines. After insertion of the form of multidimensional
transition matrix $W$ in Eq. (31) above, and performing summations, we obtain Eq. (1)
in the main text.

By using the above pattern of transition probabilities we assume that at sufficiently
short time step only one transition can take place, i.e., simultaneous transitions
in different spines are much less likely, and thus can be neglected. Indeed, since the
local basic transitions between mesoscopic states in individual spines are of the order
of several minutes (Urban et al 2020; Table 1), the likelihood that two or more such
transitions in two or more spine take place simultaneously in a short period of time
(much smaller than a minute) is small. This type of locality of explicit synaptic interactions
allows us to analyze the dynamics of global system of $N$ interacting spines.

%\vspace{1.3cm}
\newpage

\noindent{\bf\large Appendix B: Validity of the ``pair approximation'' for
analytically solvable model.}

In this section we provide conditions that much be satisfied for applying the pair
approximation in a case that can be treated analytically, which is a simplified
Ising model in thermal equilibrium. We consider 3 interacting units ($i=1,2,3$)
forming a linear ordered chain, similar as in Fig. 1, but each unit having only two
states $s_{i}= -1$ or $s_{i}= 1$. Additionally, in this model, nearest neighbors
interact strongly with the coupling $J$, whereas remote units (i.e. 1 and 3) interact
weakly with the coupling $\kappa$, which is much smaller than $J$. Our goal is to
check how accurate is the pair approximation as we increase the strength of remote
coupling $\kappa$ in relation to $J$.

The equilibrium probability of finding a given configuration of units $s_{1}, s_{2}, s_{3}$
has the form (Feynman 1972)

\begin{eqnarray}
  P(s_{1},s_{2},s_{3})= Z^{-1} e^{-J(s_{1}s_{2}+s_{2}s_{3})-\kappa s_{1}s_{3}}
\end{eqnarray}\\
where $Z^{-1}$ is the normalization factor.
The two-point marginal probabilities are given by

\begin{eqnarray}
  P(s_{1},s_{2})= 2Z^{-1} e^{-Js_{1}s_{2}} \cosh(Js_{2}+\kappa s_{1}),
\end{eqnarray}
\begin{eqnarray}
  P(s_{2},s_{3})= 2Z^{-1} e^{-Js_{2}s_{3}} \cosh(Js_{2}+\kappa s_{3}).
\end{eqnarray}\\
The one-point marginal probability for the middle unit is

\begin{eqnarray}
  P(s_{2})= 2Z^{-1}\big[ e^{Js_{2}} \cosh(Js_{2}-\kappa)
    + e^{-Js_{2}} \cosh(Js_{2}+\kappa) \big].
\end{eqnarray}\\

These four probabilities is all we need to quantify the accuracy
of the pair approximation, which in our case is represented by 
$P(s_{1},s_{2},s_{3})\approx P(s_{1},s_{2})P(s_{2},s_{3})/P(s_{2})$.
The numerical accuracy of this approximation can be assessed by
defining the ratio $R_{3}$ as

\begin{eqnarray}
R_{3}\equiv \frac{P(s_{1},s_{2})P(s_{2},s_{3})}{P(s_{2})P(s_{1},s_{2},s_{3})},
\end{eqnarray}\\
and looking how much $R_{3}$ deviates from unity.
Since $R_{3}$ depends on configurations $s_{1},s_{2},s_{3}$, it is good to
determine the mean value of $R_{3}$, averaged over all these states,
i.e., $\langle R_{3}\rangle= \frac{1}{2^{3}} \sum_{s_{1},s_{2},s_{3}} R_{3}$.

After some straightforward algebra, we can find $\langle R_{3}\rangle$ as

\begin{eqnarray}
 \langle R_{3}\rangle = \frac{ e^{\kappa}\big[1+\frac{1}{4}(e^{-4\kappa}-1)\big]
    + e^{-\kappa}\big[1+\frac{1}{4}(e^{4\kappa}-1)\big]\cosh(2J) }
  {e^{\kappa} + e^{-\kappa}\cosh(2J)}.
\end{eqnarray}\\
From this formula it is clear that for $\kappa\mapsto 0$ we get
$\langle R_{3}\rangle \mapsto 1$, regardless of the value of $J$.
For large coupling $J$ ($J \gg 1$), we obtain
$\langle R_{3}\rangle \approx 1+\frac{1}{4}(e^{4\kappa}-1)$,
which means that $\langle R_{3}\rangle$ is essentially close to 1 for
small $\kappa$. For example, for $\kappa=0.1$ we get 
$\langle R_{3}\rangle=1.12$,  for $\kappa=0.3$ we get 
$\langle R_{3}\rangle=1.58$, and higher values of $\kappa$ increase
$\langle R_{3}\rangle$ even further, which breaks the pair approximation.
For intermediate values of $J$, e.g. $J=1$, the ratio $\langle R_{3}\rangle$
achieves value 1.55 for $\kappa=0.4$, which is a slightly larger value
than for the strong coupling case.

\newpage
%\vspace{1.1cm}

\noindent{\bf\large Supplementary Information}

The code for performed computations is provided in the Supplementary Material.

\vspace{1.1cm}

\noindent{\bf\large Acknowledgments}

The work was supported by the Polish National Science Centre (NCN) grant number
2021/41/B/ST3/04300 (JK).

\vspace{1.5cm}

%\noindent{\bf\large Competing interests}
%
%There is no competing interests to declare.

%\vspace{1cm}

%\noindent{\bf\large Ethic statement}
%
%Ethic statements does not apply to this study, because all experimental data were
%collected from other sources.

\newpage

\vspace{1.5cm}

\noindent{\bf\large  References} \\
%\begin{thebibliography}{99}
%\bibitem{attwell}
Attwell D, Laughlin SB (2001) An energy budget for signaling in the 
gray matter of the brain. {\it J. Cereb. Blood Flow Metabol.} 
{\bf 21}: 1133-1145.  \\
%\bibitem{balasubramanian}
Balasubramanian V, Kimber D, Berry MJ (2001) Metabolically efficient
information processing. {\it Neural. Comput.} {\bf 13}: 799-815.  \\
%\bibitem{barrett}
Barrett AB, Billings GO, Morris RGM, van Rossum MCW (2009) State based model
of long-term potentiation and synaptic tagging and capture.
{\it PLoS Comput. Biol.} {\bf 5}: e1000259.  \\
%\bibitem{basu}
Basu S, Saha PK, Roszkowska M, Magnowska M, Baczynska E, et al (2018) Quantitative 3-D
morphometric analysis of individual dendritic spines. {\it Sci. Rep.} {\bf 8}: 3545.  \\
%\bibitem{benna}
Benna MK, Fusi S (2016) Computational principles of synaptic memory
consolidation. {\it Nature Neurosci.} {\bf 19}: 1697-1706.    \\
%\bibitem{bennett1982}
Bennett CH (1982) The thermodynamics of computation - a review. 
{\it Int. J. Theor. Physics} {\bf 21}: 905-940.  \\
%\bibitem{berut}
Berut A, Arakelyan A, Petrosyan A, Ciliberto S, Dillenschneider R, Lutz E
(2012) Experimental verification of Landauer's principle linking information
and thermodynamics. {\it Nature} {\bf 483}: 187-190.  \\
%\bibitem{bokota}
Bokota G, Magnowska M, Kusmierczyk T, Lukasik M, et al (2016) Computational approach
to dendritic spine taxonomy and shape transition analysis. {\it Front. Comput. Neurosci.}
{\bf 10}: 140.  \\
%\bibitem{bonhoeffer}
Bonhoeffer T, Yuste R (2002) Spine motility: phenomenology, mechanisms,
and function. {\it Neuron} {\bf 35}: 1019-1027.   \\
%\bibitem{bourne}
Bourne JN, Harris KM (2008) Balancing structure and function at hippocampal
dendritic spines. {\it Annual Rev. Neurosci.} {\bf 31}: 47-67.  \\
%\bibitem{chaudhuri}
Chaudhuri R, Fiete I (2016) Computational principles of memory.
{\it Nature Neurosci.} {\bf 19}: 394-403.   \\
%\bibitem{choquet}
Choquet D, Triller A (2013) The dynamic synapse. {\it Neuron} {\bf 80}: 691-703.
\\
%\bibitem{feynman}
Feynman RP (1972). {\it Statistical Mechanics: A set of lectures.}
Westview Press.  \\
%\bibitem{fusi2005}
Fusi S, Drew PJ, Abbott LF (2005) Cascade models of synaptically stored
memories. {\it Neuron} {\bf 45}: 599-611.  \\
%\bibitem{gardiner}
Gardiner CW (2004) {\it Handbook of Stochastic Methods.} Berlin: Springer. \\
Glauber RJ (1963) Time-dependent statistics of the Ising model. {\it J. Math.
  Phys.} {\bf 4}: 294-307.  \\
Govindarajan A, Kelleher RJ, Tonegawa S (2006) A clustered plasticity model of
long-term memory engrams. {\it Nat. Rev. Neurosci.} {\bf 7}: 575-583.  \\
%\bibitem{holtmaat}
Holtmaat AJ, Trachtenberg JT, Wilbrecht L, Shepherd GM, Zhang X, et al 
(2005) Transient and persistent dendritic spines in the neocortex in vivo.
{\it Neuron} {\bf 45}: 279-291.   \\
%\bibitem{kandel}
Kandel ER, Dudai Y, Mayford MR (2014) The molecular and systems biology
of memory. {\it Cell} {\bf 157}: 163-186.   \\
%\bibitem{karbowski2009}
Karbowski J (2009) Thermodynamic constraints on neural dimensions, firing rates,
brain temperature and size. {\it J. Comput. Neurosci.} {\bf 27}: 415-436.  \\
%\bibitem{karbowski2012}
Karbowski J (2012) Approximate invariance of metabolic energy per synapse
during development in mammalian brains. {\it PLoS ONE} {\bf 7}: e33425.   \\
%\bibitem{karbowski2019}
Karbowski J (2019) Metabolic constraints on synaptic learning and memory.
{\it J. Neurophysiol.} {\bf 122}: 1473-1490.  \\
%\bibitem{karbowski2021}
Karbowski J (2021) Energetics of stochastic BCM type synaptic plasticity and
storing of accurate information. {\it J. Comput. Neurosci.} {\bf 49}: 71-106.  \\
Karbowski J, Urban P (2022) Information encoded in volumes and areas of
dendritic spines is nearly maximal across mammalian brains.
(www.biorxiv/2021.12.30.474505) Preprint.  \\ 
%\bibitem{kasai}
Kasai H, Matsuzaki M, Noguchi J, Yasumatsu N, Nakahara H (2003) 
Structure-stability-function relationships of dendritic spines.
{\it Trends Neurosci.} {\bf 26}: 360-368.   \\
Kastellakis G, Poirazi P (2019) Synaptic clustering and memory formation.
{\it Front. Mol. Neurosci.} {\bf 12}: 300.  \\
%\bibitem{kennedy}
Kennedy MB (2000) Signal-processing machines at the postsynaptic density.
  {\it Science} {\bf 290}: 750-754.   \\
%\bibitem{laughlin1998}
Laughlin SB, de Ruyter van Steveninck RR, Anderson JC (1998) The metabolic
cost of neural information. {\it Nature Neurosci.} {\bf 1}: 36-40.  \\
%\bibitem{maxwell_demon}
Leff HS, Rex AF (1990) {\it Maxwell's Demon: Entropy, Information, Computing}.
Princeton Univ. Press: Princeton, NJ. \\
%\bibitem{leibold}
  Leibold C, Kempter R (2008) Sparseness constrains the prolongation of
  memory lifetime via synaptic metaplasticity. {\it Cereb. Cortex} {\bf 18}:
  67-77.   \\
%\bibitem{levy}
Levy WB, Baxter RA (1996) Energy efficient neural codes. {\it Neural Comput.} 
{\bf 8}: 531-543.   \\
%\bibitem{levy1999}
Levy WB, Baxter RA (2002) Energy-efficient neuronal computation via
quantal synaptic failures . {\it J. Neurosci.} 
{\bf 22}: 4746-4755.   \\
%\bibitem{levy2021}
Levy WB, Calvert VG (2021) Communication consumes 35 times more energy than
computation in the human cortex, but both costs are needed to predict synapse
number. {\it Proc. Natl. Acad. Sci. USA} {\bf 118}: e2008173118.  \\
%\bibitem{li}
Li HL, van Rossum MCW (2020) Energy efficient synaptic plasticity. {\it eLife}
{\bf 9}: e50804.   \\
%\bibitem{loewenstein}
Loewenstein Y, Kuras A, Rumpel S (2011) Multiplicative dynamics underlie
the emergence of the log-normal distribution of spine sizes in the 
neocortex in vivo. {\it J. Neurosci.} {\bf 31}: 9481-9488.   \\
Maes C, Redig F, van Moffaert A (2000) On the definition of entropy production,
via examples. {\it J. Math. Phys.} {\bf 41}: 1528.   \\
%\bibitem{makino}
Makino H, Malinow R (2011) Compartmentalized versus global synaptic plasticity
on dendrites controlled by experience. {\it Neuron} {\bf 72}: 1001-1011.  \\
Matsuda H, Ogita N, Sasaki A, Sato K (1992) Statistical mechanics of population:
The lattice Lotka-Volterra Model. {\it Prog. Theor. Physics}  {\bf 88}: 1035-1049. \\
%\bibitem{meyer}
Meyer D, Bonhoeffer T, Scheuss V (2014) Balance and stability
of synaptic structures during synaptic plasticity. {\it Neuron}
{\bf 82}: 430-443. \\
%\bibitem{miller}
Miller P, Zhabotinsky AM, Lisman JE, Wang X-J (2005) The stability of a stochastic
CaMKII switch: Dependence on the number of enzyme molecules and protein turnover.
{\it PLoS Biol.} {\bf 3}: e107.   \\
%\bibitem{montgomery}
Montgomery JM, Madison DV (2004) Discrete synaptic states define a major
mechanism of synaptic plasticity. {\it Trends Neurosci.}  {\bf 27}: 744-750.   \\
%\bibitem{nicolis}
Nicolis G, Prigogine I (1977) {\it Self-Organization in Nonequilibrium Systems}.
Wiley: New York, NY.   \\
%\bibitem{niven}
Niven B, Laughlin SB (2008) Energy limitation as a selective pressure on
the evolution of sensory systems. {\it J. Exp. Biol.} {\bf 211}: 1792-1804.  \\
%\bibitem{parrondo}
Parrondo JMR, Horowitz JM, Sagawa T (2015) Thermodynamics of information.
{\it Nature Physics} {\bf 11}: 131-139.   \\
%\bibitem{peliti}
Peliti L, Pigolotti S (2021) {\it Stochastic Thermodynamics: An Introduction.}
Princeton: Princeton Univ. Press.  \\
%\bibitem{poirazi}
Poirazi P, Mel BW (2001) Impact of active dendrites and structural plasticity
on the memory capacity of neural tissue. {\it Neuron} {\bf 29}: 779-796.  \\
%\bibitem{poo}
Poo M-m, Pignatelli M, Ryan TJ, Tonegawa S, Bonhoeffer T, Martin KC, Rudenko A,
Tsai L-H, Tsien RW, Fishell G, et al (2016) What is memory? The present state of
the engram. {\it BMC Biol.} {\bf 14}: 40.   \\
%\bibitem{rieke}
Rieke F, Warland D, de Ruyter R, Bialek W (1999) {\it Spikes: Exploring the neural code.}
Cambridge, MA: MIT Press. \\ 
%\bibitem{schnakenberg} 
Schnakenberg J (1976) Network theory of microscopic and macroscopic behavior
of master equation systems. {\it Reviews of Modern Physics} {\bf 48}: 571-585.  \\
%\bibitem{seifert}
Seifert U (2012) Stochastic thermodynamics, fluctuation theorems and molecular machines.
{\it Rep. Prog. Phys.}  {\bf 75}: 126001.  \\
%\bibitem{sheng}
Sheng M, Hoogenraad CC (2007) The postsynaptic architecture of excitatory
synapses: a more quantitative view. {\it Annu. Rev. Biochem.} {\bf 76}:
823-847.  \\
%\bibitem{statman}
Statman A, Kaufman M, Minerbi A, Ziv NE, Brenner N (2014) 
Synaptic size dynamics as an effective stochastic process.
{\it PLoS Comput. Biol.} {\bf 10}: e1003846.   \\
%\bibitem{takeuchi}
Takeuchi T, Duszkiewicz AJ, Morris RGM (2014) The synaptic plasticity and memory
hypothesis: encoding, storage and persistence.  {\it Phil. Trans. R. Soc. B}
{\bf 369}: 20130288.   \\
%\bibitem{urban}  
Urban P, Tabar VR, Denkiewicz M, Bokota G, Das N, et al (2020) The mixture of
autoregressive hidden Markov models of morphology for dendritic spines during
activation process. {\it J. Comput. Biol.}  {\bf 27}: 1471-1485.   \\
%\bibitem{vanbaalen}  
Van Baalen M (2000) Pair approximation for different spatial geometries.
In: {\it The Geometry of Ecological Interactions: Simplifying Spatial Complexity.}
Eds: Dieckmann U et al, pp. 359-387. Cambridge Univ. Press.   \\
%\bibitem{vandenbroeck}  
  Van den Broeck C, Esposito M (2015) Ensemble and trajectory thermodynamics:
  A brief introduction. {\it Physica A} {\bf 418}: 6-16.   \\
%\bibitem{volgushev}
Volgushev M, et al (2004) Probability of transmitter release at neocortical
 synapses at different temperatures. {\it J. Neurophysiol.}  {\bf 92}: 212-220.  \\
%\bibitem{winnubst}
 Winnubst J, Lohmann C, Jontes J, Wang H, Niell C (2012) Synaptic clustering
 during development and learning: the why, when, and how.  {\it Front. Mol. Neurosci.}
 {\bf 5}: 70.   \\
%\bibitem{yadav}
Yadav A, Gao YZ, Rodriguez A, Dickstein DL, et al (2012) Morphologic evidence
for spatially clustered spines in apical dendrites of monkey neocortical
pyramidal cells. {\it J. Comp. Neurol.} {\bf 520}: 2888-2902.   \\
%\bibitem{yang}
  Yang G, Pan F, Gan W-B (2009) Stably maintained dendritic spines are associated
  with lifelong memories. {\it Nature}  {\bf 462}: 920-924.   \\
%\bibitem{yasumatsu}
Yasumatsu N, Matsuzaki M, Miyazaki T, Noguchi J, Kasai H (2008)
Principles of long-term dynamics of dendritic spines. {\it J. Neurosci.}
{\bf 28}: 13592-13608.   \\

%\bibitem{bartol}
%Bartol TM, Bromer C, Kinney J, Chirillo MA, Bourne JN, Harris KM, Sejnowski TJ
%(2015) Nanoconnectomic upper bound on the variability of synaptic plasticity.
%{\it eLife} {\bf 4}: e10778.  \\
%\bibitem{benavides2013}
%Benavides-Piccione R, Fernaud-Espinosa I, Robles V, Yuste R, DeFelipe J
%(2013) Age-based comparison of human dendritic spine structure using
%complete three-dimensional reconstructions. {\it Cerebral Cortex}
%{\bf 23}: 1798-1810.  \\

%\bibitem{karbowski2007}
%Karbowski J (2007) Global and regional brain metabolic scaling and its
%functional consequences. {\it BMC Biology} {\bf 5}: 18.  \\
%\bibitem{karbowski2012}
%Karbowski J (2014) Constancy and trade-offs in the neuroanatomical and metabolic
%design of the cerebral cortex. {\it Front. Neural Circuits} {\bf 8}: 9.  \\
%\bibitem{karbowski2015}
%Karbowski J (2015) Cortical composition hierarchy driven by spine proportion
%economical maximization or wire volume minimization.  {\it PloS Comput. Biol. } 
%{\bf 11}: e1004532.  \\
%\bibitem{press}
%Press WH, Teukolsky SA, Vetterling WT, Flannery BP (1992) {\it Numerical
%Recipes in Fortran.} Cambridge: Cambridge Univ. Press.
%

%\end{thebibliography}

%\vspace{1.5cm}

\newpage

{\bf \large Figure Captions}

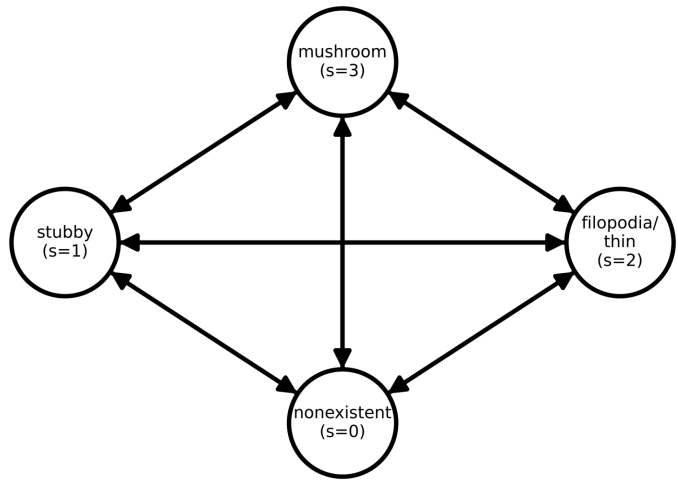
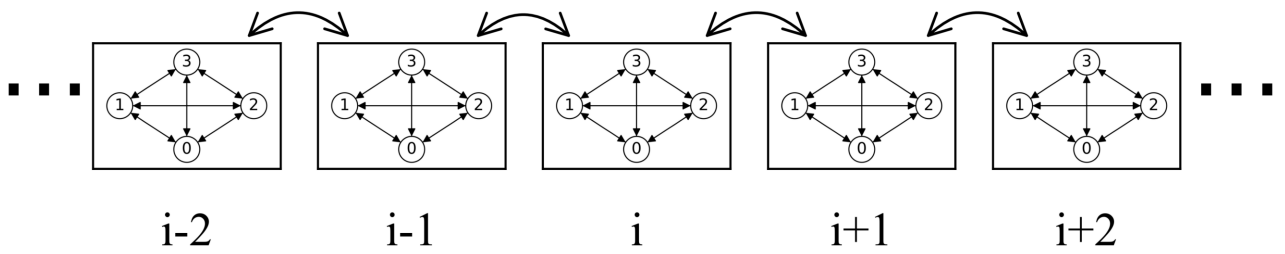
{\bf Fig. 1} \\
{\bf Mesoscopic model of dendritic spines and their interactions on a dendrite.} 
A) Four state stochastic model of a dendritic spine with transitions between the
mesoscopic states. B) Interactions between spines on a dendrite are determined only
by nearest neighbors, in such a way that spine's intrinsic transition rates depend
on the states of neighboring spines. These interactions can be thought as representing
inflow and outflow of different molecules between spines.

\vspace{0.3cm}

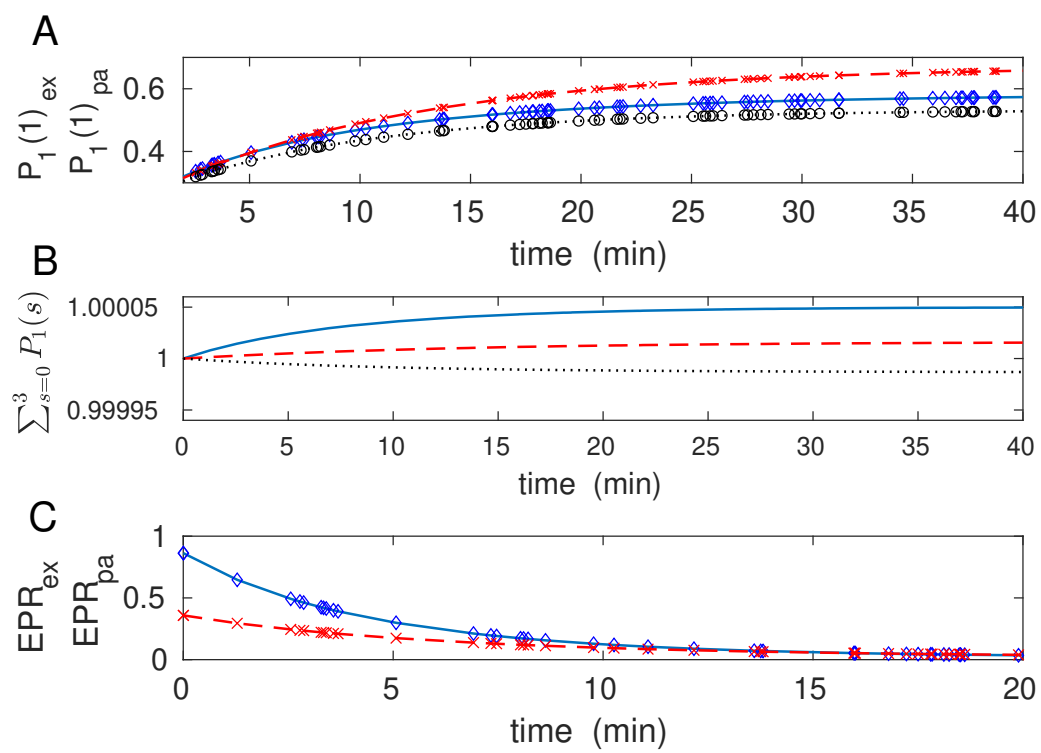
{\bf Fig. 2} \\
{\bf Comparison of an exact solution and pair approximation for N=4 interacting spines.} 
A) Time dependence of exact $P_{1}(1)_{ex}$ and approximated $P_{1}(1)_{pa}$
probability $P(s_{1}=1)$ for three different couplings $\gamma$.
Both probabilities are essentially indistinguishable.
Exact solutions correspond to solid ($\gamma=-0.9$),
dashed ($\gamma=0.1$), and dotted ($\gamma=0.9$) lines.
Pair approximations correspond to diamonds ($\gamma=-0.9$), x ($\gamma=0.1$),
and circles ($\gamma=0.9$).
B) Normalization condition for probabilities of spine no 1, i.e. $\sum_{s_{1}=0}^{3} P(s_{1})$.
Solid line for $\gamma=-0.9$, dashed line for $\gamma=0.1$, and dotted line for
$\gamma=0.9$. Note that the sum deviates from unity by a very small number less than
$5\cdot 10^{-5}$.
C) Time dependence of exact (EPR$_{ex}$) and approximated (EPR$_{pa}$) entropy
production rate. Exact EPR$_{ex}$ correspond to solid ($\gamma=-0.9$) and
dashed ($\gamma=0.1$) lines.
Pair approximations EPR$_{pa}$ correspond to diamonds ($\gamma=-0.9$) and
x ($\gamma=0.1$). Note an excellent matching of EPR$_{pa}$ to EPR$_{ex}$.

\vspace{0.3cm}

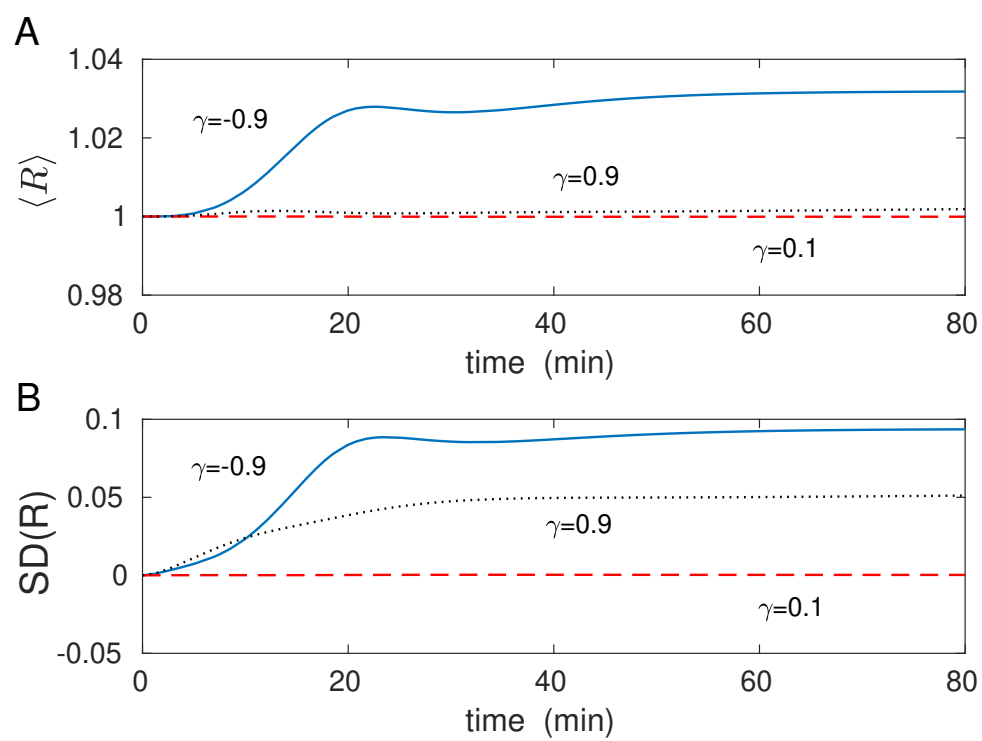
{\bf Fig. 3} \\
{\bf Accuracy measure for pair approximation with N=4 interacting spines.} 
A) Average ratio $\langle R\rangle$ as a function of time for different magnitudes
of coupling between the spines.
B) Similar as in A but for standard deviation of the ratio $R$. Note that SD(R) is
for moderate values of $\gamma$ much less than 0.05, and maximally achieves the value
$\sim 0.1$ in the extreme case $\gamma \mapsto -1$.

\vspace{0.3cm}

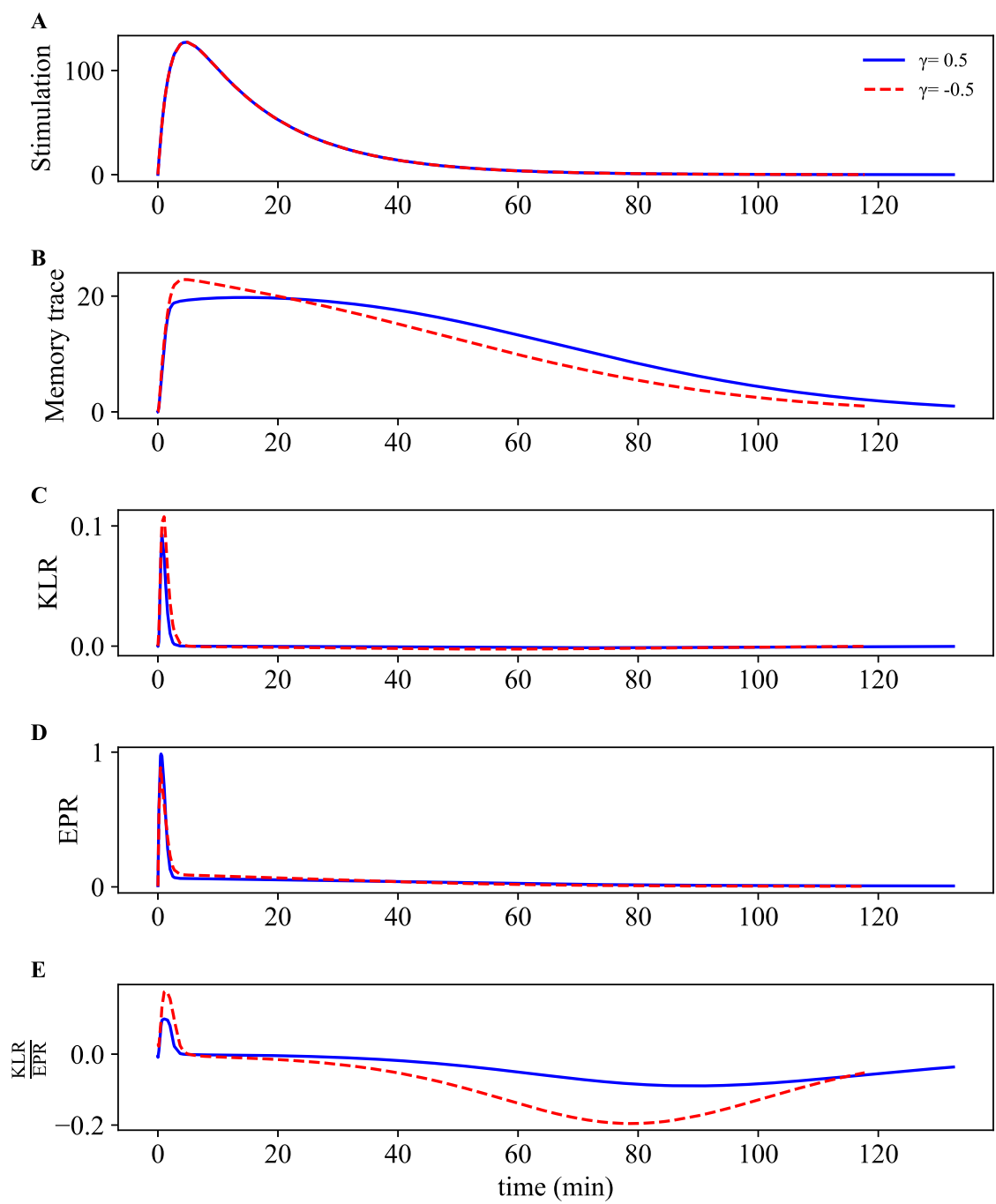
{\bf Fig. 4} \\
{\bf Dynamics of memory trace, and the rates of information gain and energy.} 
Temporal dependence of A) stimulation function $f(t)$ (amplifier of transitions between
the states), B) memory trace MT, C) information gain rate per spine (Kullback-Leibler
divergence rate, KLR/$N$), D) entropy production rate (energy rate) per spine (EPR/$N$ in units
of $\epsilon/min$), and of E) the ratio of information gain rate to entropy production rate
per spine (KLR/EPR). Learning phase, equivalent to stimulation phase, lasts up to 20 min (A).
Memory phase, quantified by memory trace, starts after the end of stimulation and lasts
up to $\sim$ 120 min (B). Note that upon stimulation the rates of information gain and entropy
production (KLR and EPR) achieve peaks very fast, but they also decay very fast. In contrast,
memory trace lasts much longer. EPR value at its peak is $\sim$ $\epsilon$/min
$\approx 4.6\cdot 10^{5}$ kT/min per spine, which is about 2-3 orders of magnitude larger
than at baseline (before or long after the stimulation).

\vspace{0.3cm}

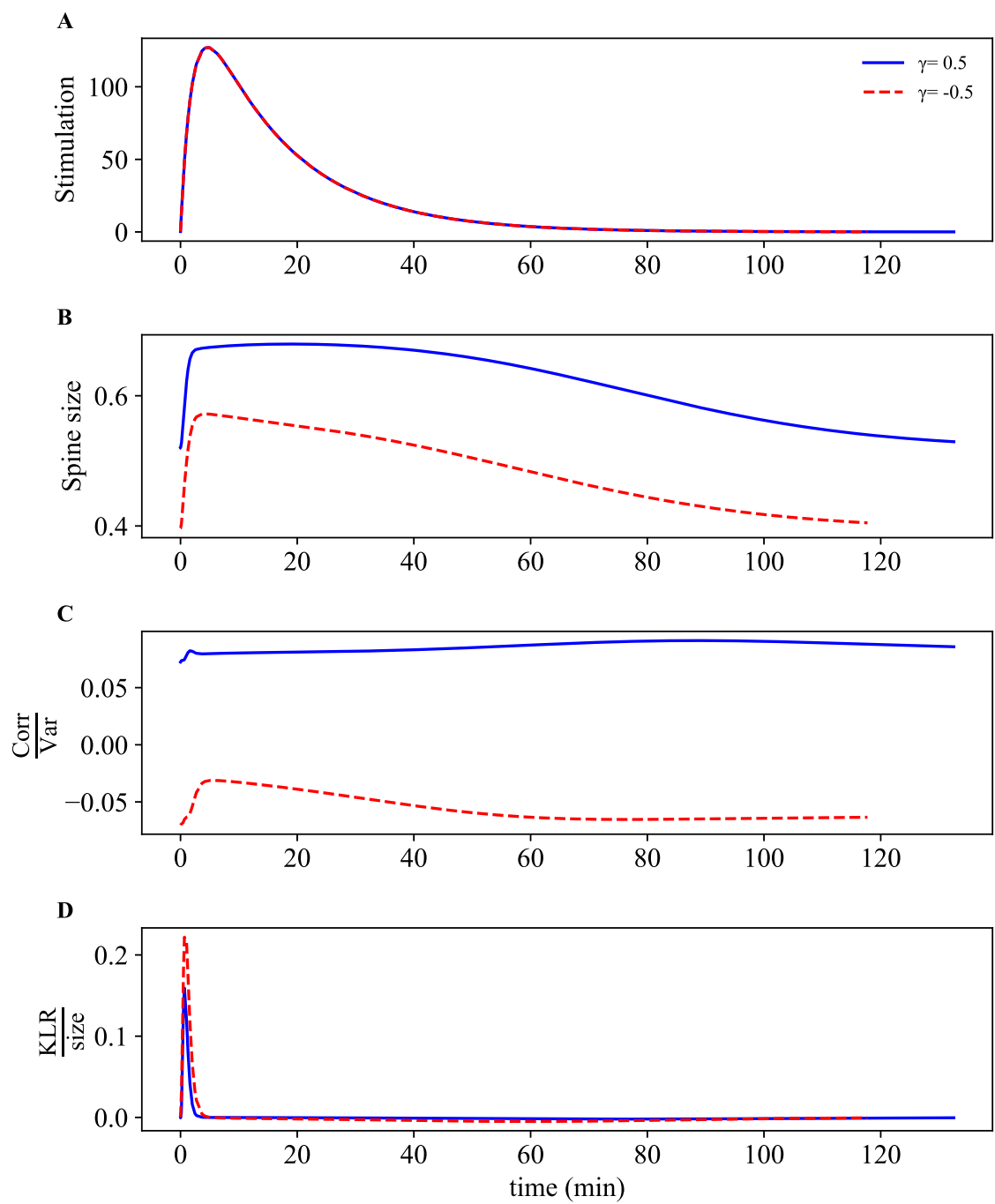
{\bf Fig. 5} \\
{\bf Dynamics of spine size, correlations, and of information gain per spine size.} 
Temporal dependence of A) stimulation function $f(t)$, B) average spine size $\langle S\rangle$
C) normalized correlations between spines, and of D) the ratio of KLR per spine to average
spine size $\langle S\rangle$.

\vspace{0.3cm}

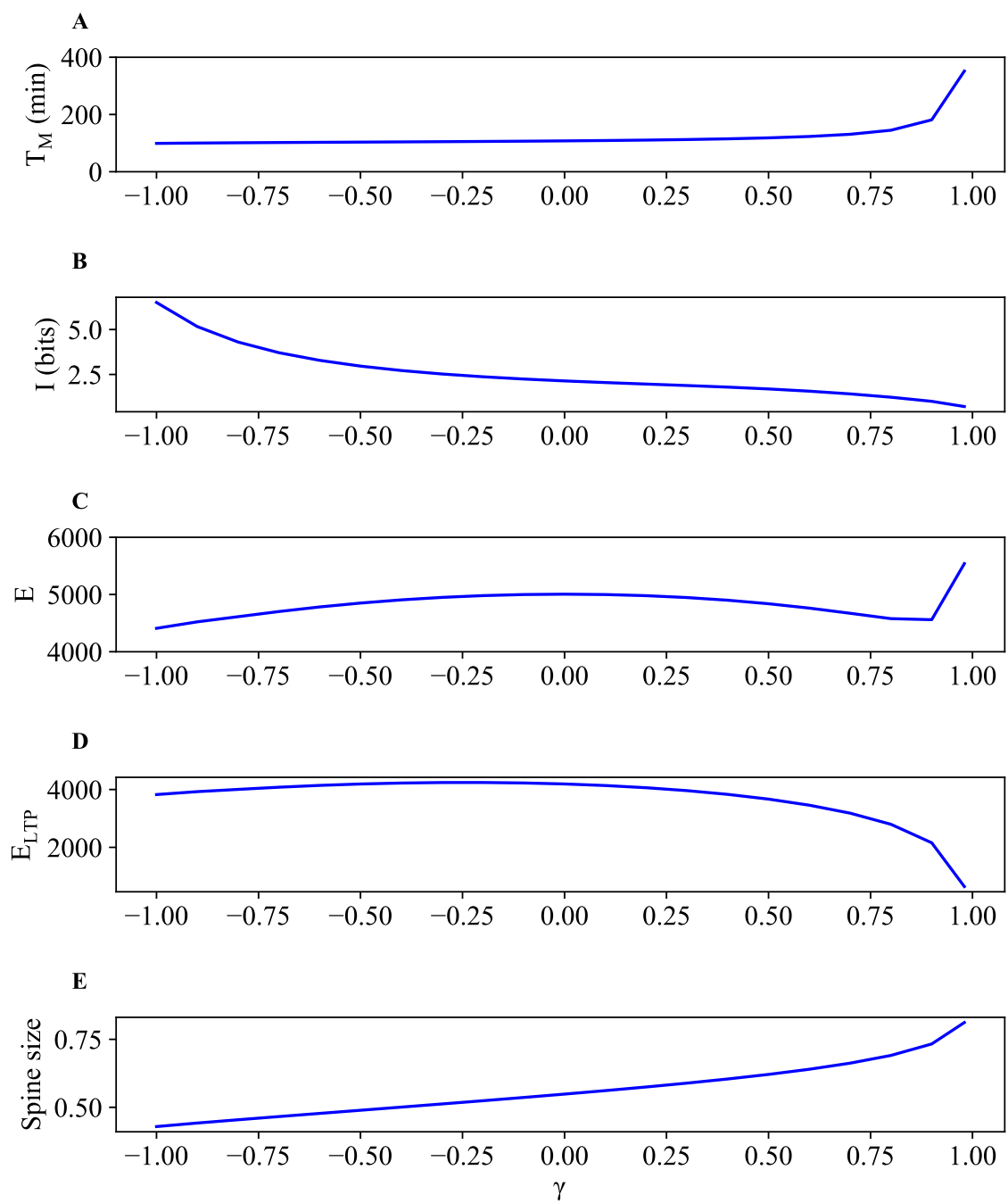
{\bf Fig. 6} \\
{\bf Dependence of memory time, information gain, energy cost, and spine size
on the synaptic cooperativity. } 
A) and B) Monotonic but opposite dependence of memory time $T_{M}$ and total information
gain $I$ on cooperativity $\gamma$. C) and D) Total energy consumption of all spines $E$
during LTP and its part $E_{ltp}$, related solely to LTP induction and maintenance, exhibit
nonmonotonic dependence on $\gamma$ (energy units are in $\epsilon$; see Methods). During
whole LTP, a typical spine used about $5\epsilon\approx 2.3\cdot 10^{6}$ kT of energy.
E) Average spine size increases monotonically with $\gamma$.

\vspace{0.3cm}

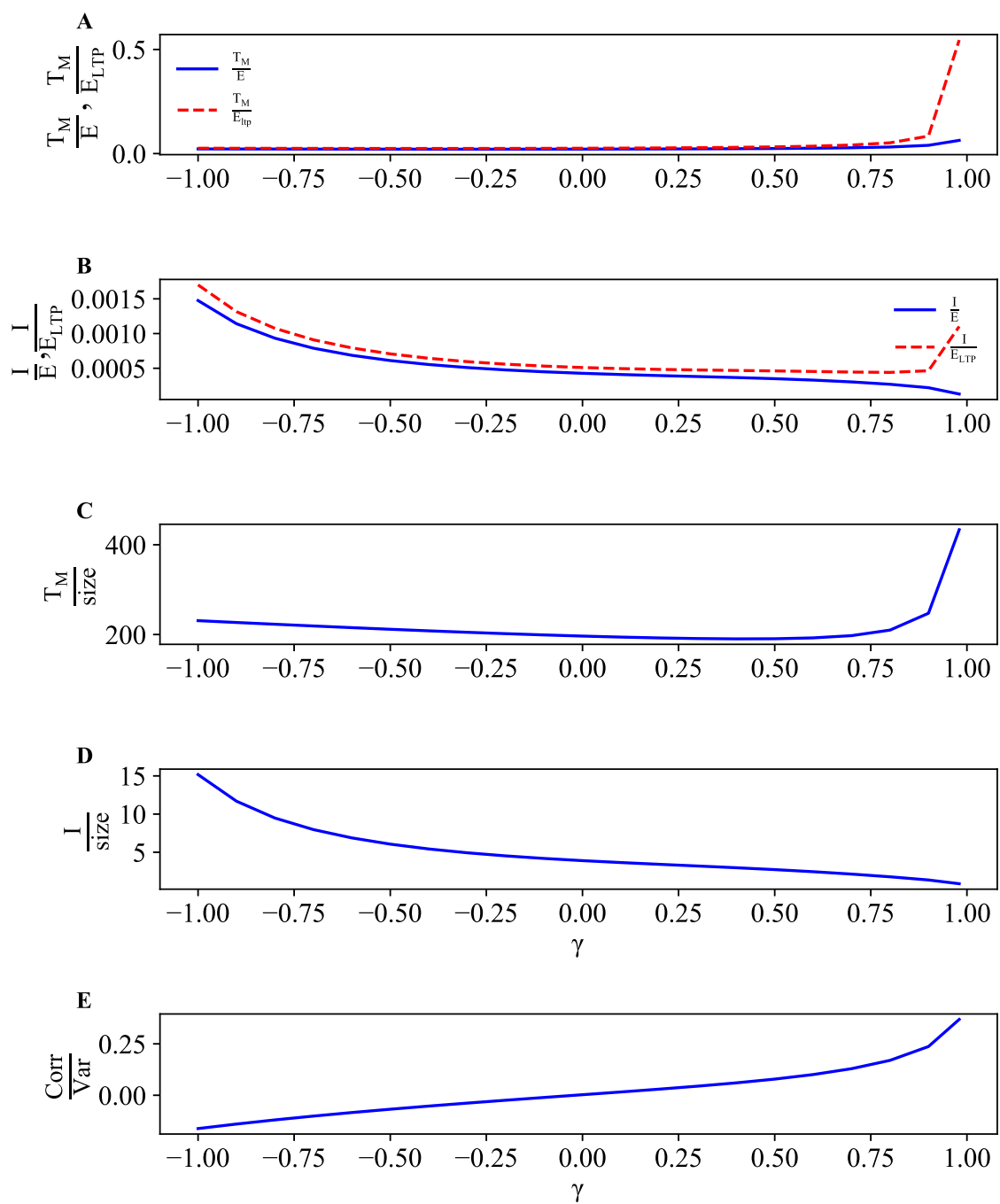
{\bf Fig. 7} \\
{\bf Energetic and structural efficiencies of memory time and information gain
  in comparison to spine correlations as functions of synaptic cooperativity. } 
Ratios of A) memory time $T_{M}$ and B) information gain $I$ to two energies $E$ and
$E_{ltp}$ exhibit two opposite dependence. Similar behavior for the ratios C) memory
time and D) information gain to average spine size $\langle S\rangle$ as functions of
$\gamma$. E) Normalized correlations always increase with $\gamma$, reaching $\sim 0.3$
for $\gamma \mapsto 1$.

\vspace{0.3cm}

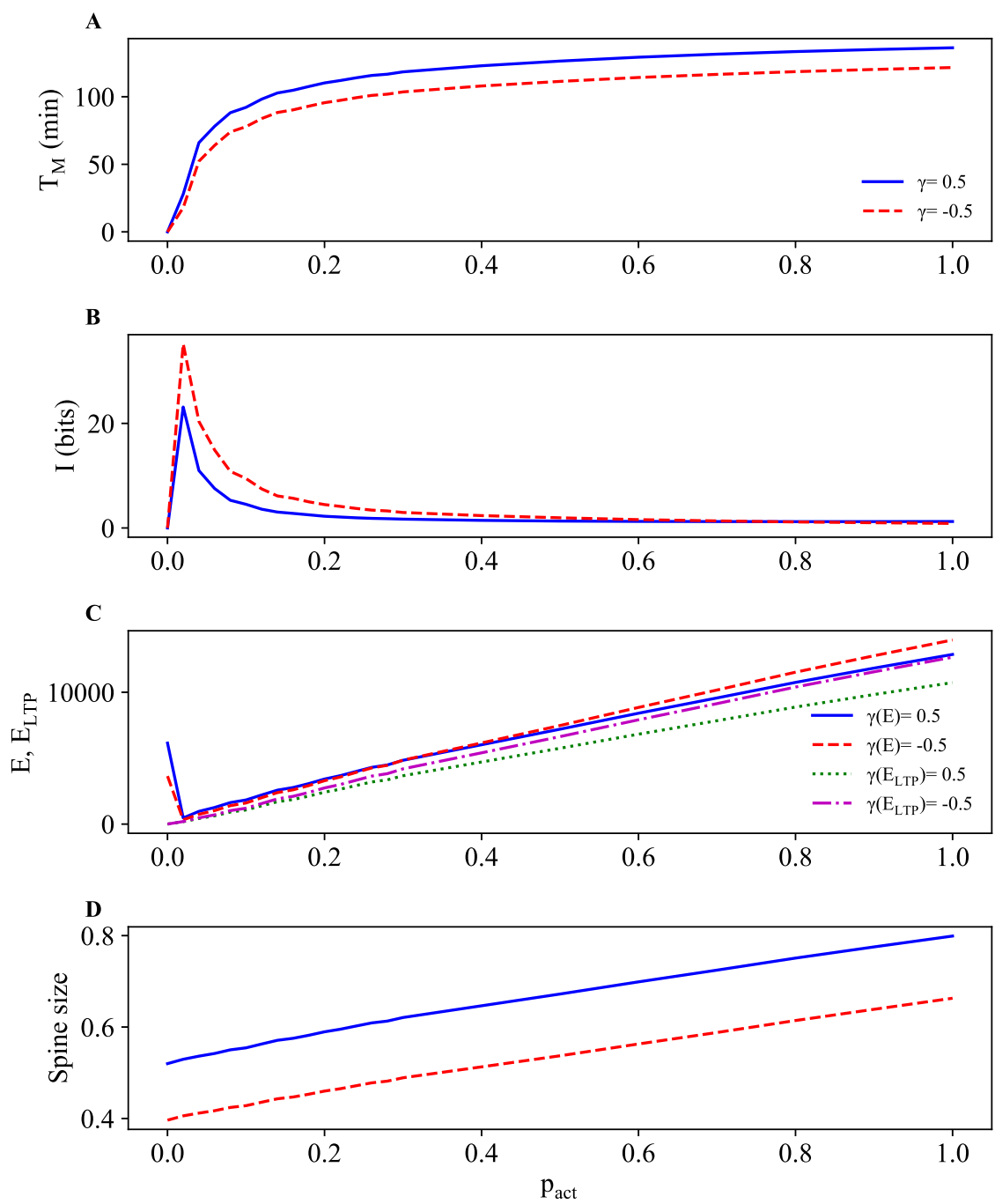
{\bf Fig. 8} \\
{\bf Dependence of memory time, information gain, energy cost, and spine size
on probability of synaptic stimulation. } 
A) Memory lifetime $T_{M}$ saturates for large $p_{act}$. B) Information gain
exhibits a sharp peak for very small $p_{act}$. C, D) Energy solely due to LTP
and average spine size increase linearly with $p_{act}$.

\vspace{0.3cm}

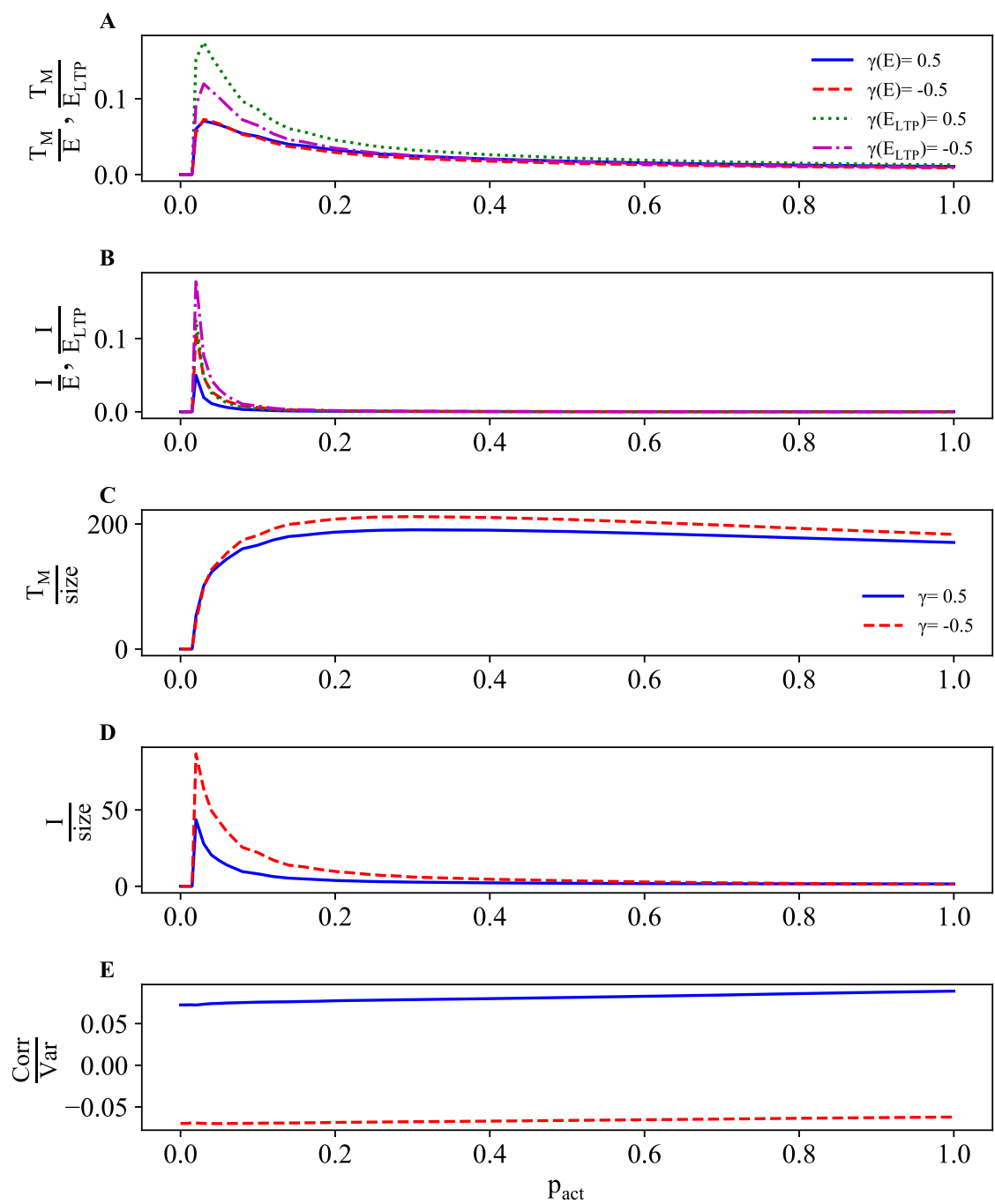
{\bf Fig. 9} \\
{\bf Energetic and structural efficiencies of memory time and information gain
as functions of probability of synaptic stimulation. } 
All the ratios of memory time and information gain to energies and to spine size
$\langle S\rangle$ exhibit maxima. At the peak, the ratio $I/E$ is $\sim 0.05$ bit/$\epsilon$
and $I/E_{ltp}$ is $\sim 0.15$ bit/$\epsilon$, which means that 1 bit of stored
information after LTP degradation requires about $10^{7}$ kT of energy.

\vspace{0.3cm}

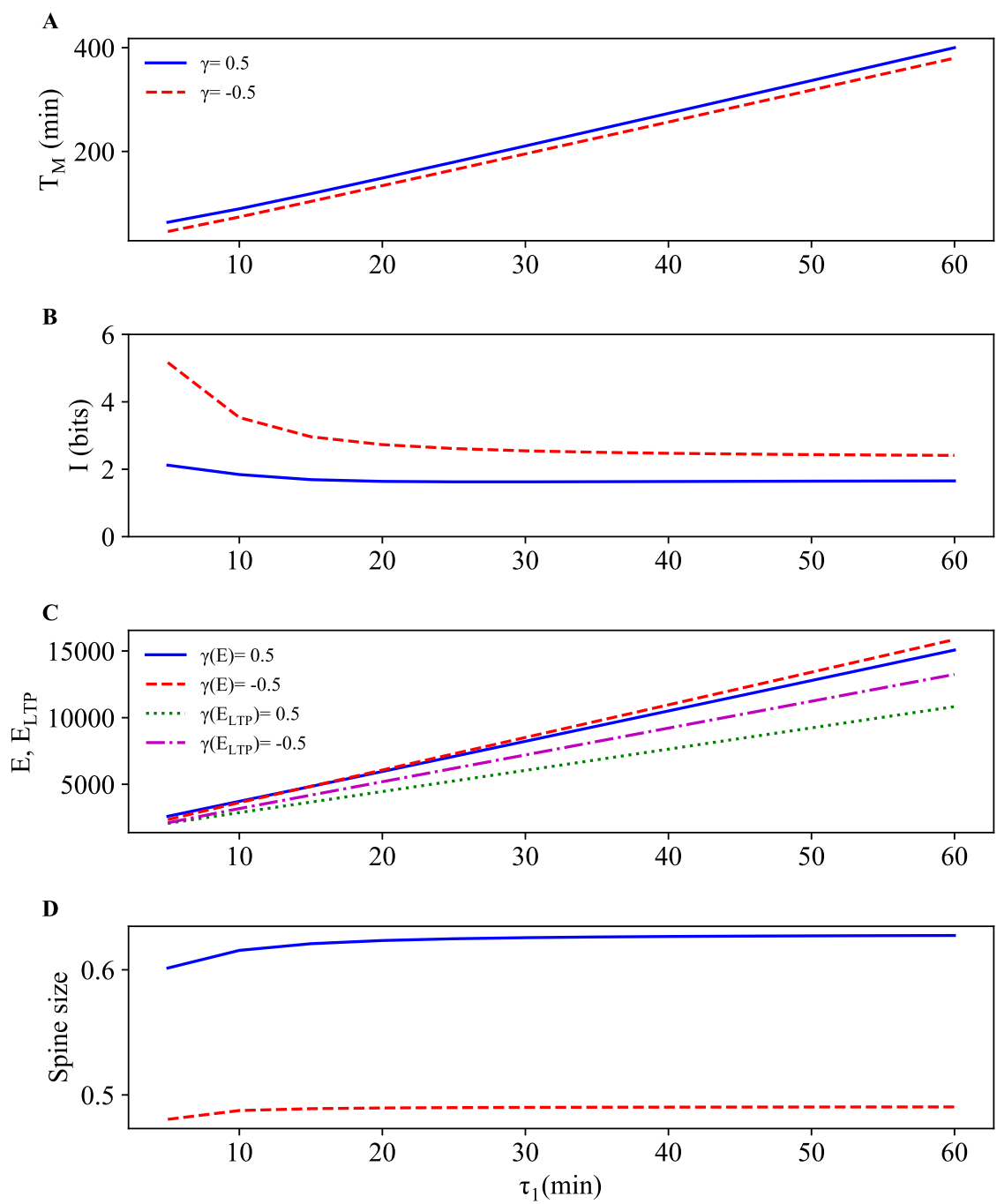
{\bf Fig. 10} \\
{\bf Dependence of memory time, information gain, energy cost, and spine size 
on duration of stimulation $\tau_{1}$. }

\vspace{0.3cm}

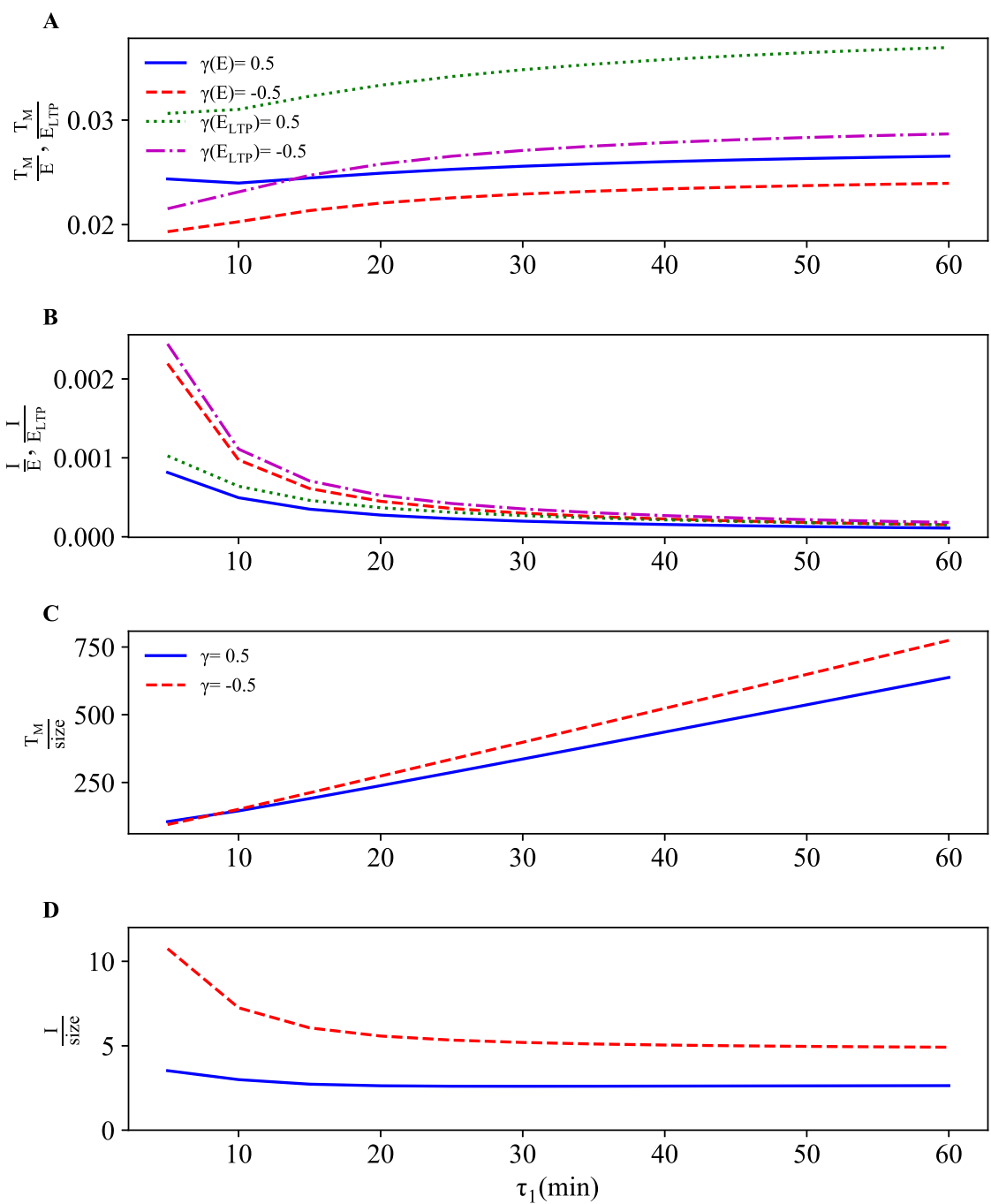
{\bf Fig. 11} \\
{\bf Energetic and structural efficiencies of memory time and information gain
as functions of duration of stimulation $\tau_{1}$. } 
Note that energetic efficiency drops with increasing the duration of stimulation
(B). Structural efficiency of memory lifetime grows linearly with $\tau_{1}$ (C).

\vspace{0.3cm}

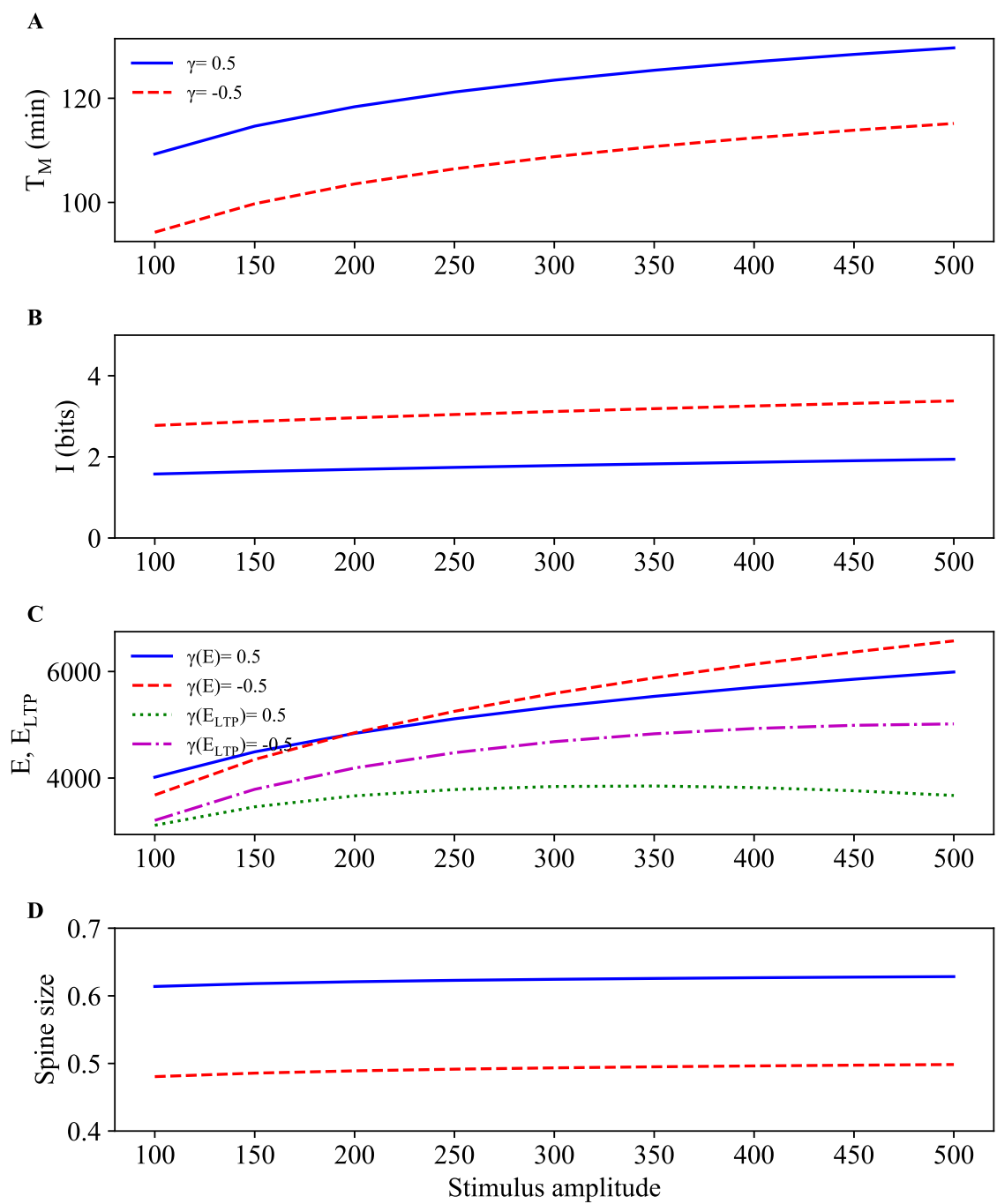
{\bf Fig. 12} \\
{\bf Dependence of memory time, information gain, energy cost, and spine size 
  on strength of stimulation A. }
Note that information gain (A) and mean spine size (D) are essentially independent
of $A$.

\vspace{0.3cm}

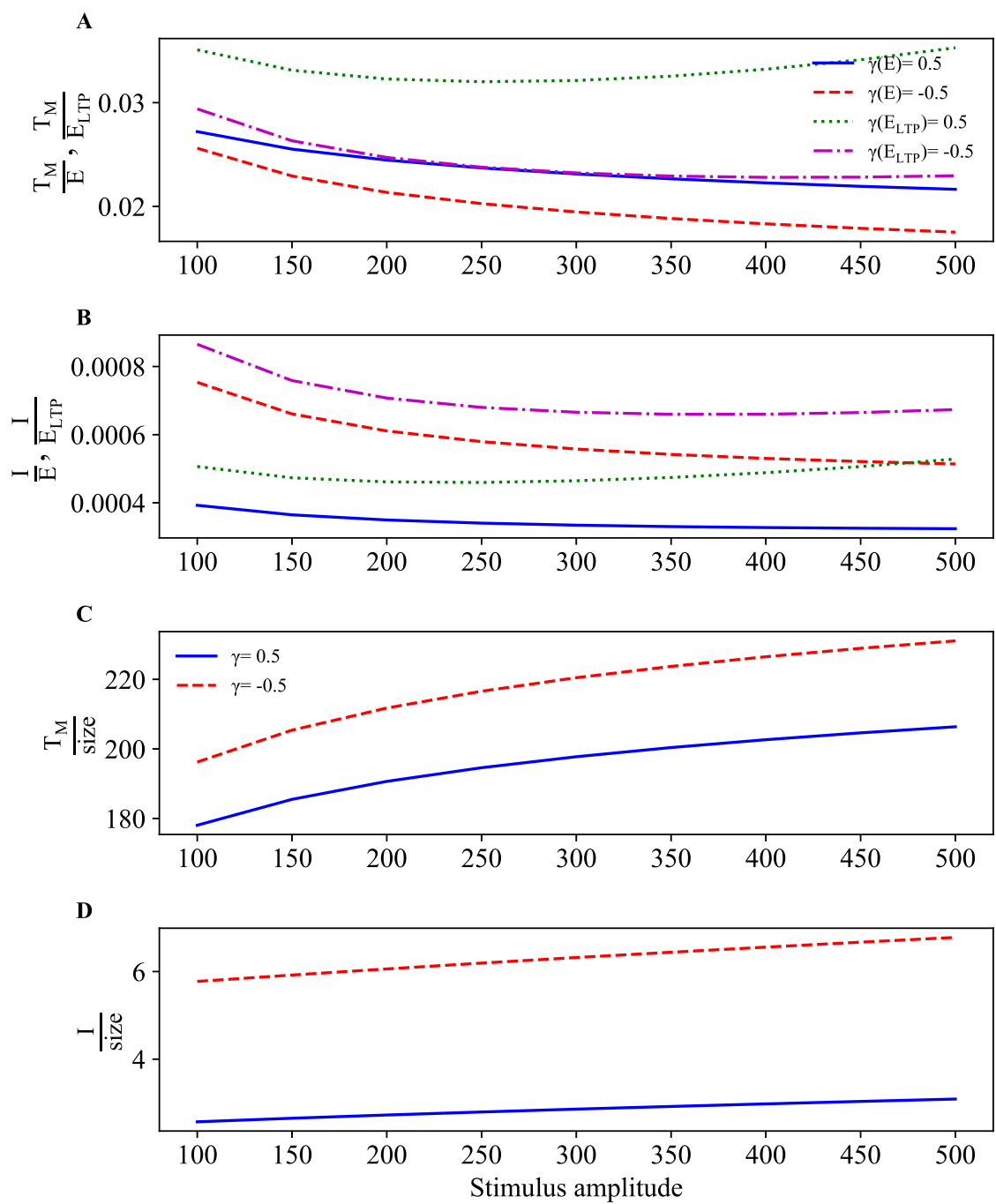
{\bf Fig. 13} \\
{\bf Energetic and structural efficiencies of memory time and information gain
as functions of strength of stimulation A. } 
Note that too large stimulation amplitudes are not beneficial for the energy
and structural efficiencies (saturation effects).

\vspace{0.3cm}

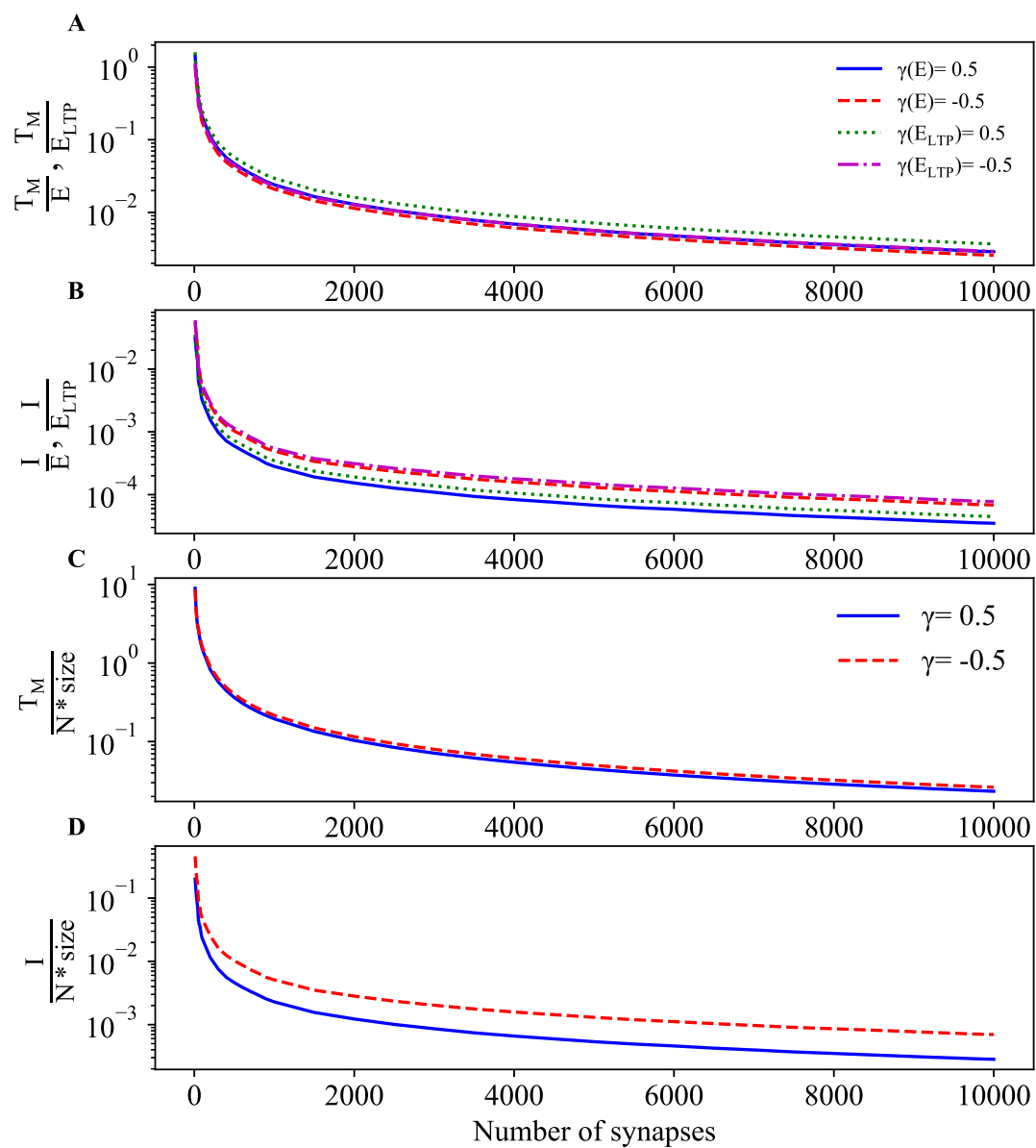
{\bf Fig. 14} \\
{\bf Energy and structural efficiencies of memory lifetime and information gain
  drop with increasing synaptic numbers $N$. } 
(A, B) Energy efficiency of $T_{M}$ and $I$ decreases dramatically, by two orders
of magnitude, with increasing the number of spines from $N=10$ to $N= 2000$, and
much slower with increasing the number of spines from $N=2000$ to $N= 10000$.
(C, D) Essentially the same declining effect is observed for structural efficiency of
$T_{M}$ and $I$, i.e., $T_{M}/(N\langle S\rangle)$ and  $I/(N\langle S\rangle)$.

\newpage

 \begin{table}[htp]
 %\label{tab:parameters}
\caption{Values of intrinsic basic transition rates $v_{s,s'}$ in each synapse.}
\begin{center}
\begin{tabular}{c | c }
Matrix element  & Value   \\
\hline

   $v_{0,1}$   &   0.019    \\
   $v_{1,0}$   &   0.065    \\
   $v_{0,2}$   &   0.015    \\
   $v_{2,0}$   &   0.007    \\
   $v_{0,3}$   &   0.022    \\
   $v_{3,0}$   &   0.001    \\
   $v_{2,1}$   &   0.003    \\
   $v_{1,2}$   &   0.061    \\
   $v_{2,3}$   &   0.009    \\
   $v_{3,2}$   &   0.015    \\
   $v_{1,3}$   &   0.049    \\
   $v_{3,1}$   &   0.008    \\

\hline

\end{tabular}
\end{center}
All diagonal elements, i.e. $v_{k,k}$, are zero.
The units are in min$^{-1}$.
\end{table}%

\newpage

 \begin{table}[htp]
 %\label{tab:parameters}
\caption{Nominal values of global parameters used in computations.}
\begin{center}
\begin{tabular}{c | c | c | c}
Description &  Variable  & Value  &  Units \\
\hline

Number of synapses  &  $N$   &  1000   &  unitless  \\

Cooperativity       &  $\gamma$  &  (-1.0,1.0)  &  unitless  \\

Probability of stimulation   &  $p_{act}$     &   0.3      &  unitless  \\

Stimulation amplitude &  $A$     &   200  &  unitless  \\

Decay time of stimulation   &   $\tau_{1}$  &   15     &  min   \\

Rising time of stimulation    &   $\tau_{2}$  &   2     &  min   \\

Energy scale for plasticity    &   $\epsilon$  &  $4.6\cdot 10^{5}$  &  kT  \\

\hline

\end{tabular}
\end{center}

Value of stimulation amplitude $A$ is motivated by experimental facts that
during LTP the rates of protein phosphorylation at PSD increase by $2-3$
orders of magnitude in respect to baseline rates (e.g. Miller et al 2005).
Energy scale $\epsilon$ for synaptic plasticity was taken from the estimate in
Karbowski (2021).

\end{table}%

%\end{narrowtext}
\end{document}